\documentclass[11pt,a4paper]{article}
\pdfoutput=1

\usepackage{jheppub}
\usepackage{amsmath,amsfonts,amssymb,amsthm}
\usepackage{url}
\usepackage{physics}
\usepackage{hyperref}
\usepackage{comment}

\numberwithin{equation}{section}

\newcommand{\fft}[2]{\frac{#1}{#2}}

\newcommand{\nn}{\nonumber}

\DeclareMathAlphabet{\mathpzc}{OT1}{pzc}{m}{it}

\allowdisplaybreaks[1]

\preprint{LITP-26-06, USTC-ICTS/PCFT-26-15}

\title{New $F^4$ invariants in five-dimensional supergravity}

\author[a]{Yide Cai,}
\author[a]{Sabarenath Jayaprakash,}
\author[a]{James T. Liu,}
\author[b,c]{Yi Pang,}
\author[b]{Robert J. Saskowski}
\emailAdd{caiyi@umich.edu, sabare@umich.edu, jimliu@umich.edu, pangyi1@tju.edu.cn, robert\_saskowski@tju.edu.cn}

\affiliation[a]{Leinweber Institute for Theoretical Physics, 
University of Michigan, Ann Arbor, MI 48109, USA}
\affiliation[b]{Center for Joint Quantum Studies and Department of Physics, School of Science,\\ Tianjin University, Tianjin 300350, China}
\affiliation[c]{Peng Huanwu Center for Fundamental Theory, Hefei, Anhui 230026, China}

\abstract{We consider four-derivative superinvariants of five-dimensional $\mathcal N=2$ supergravity coupled to $n_v\le 2$ vector multiplets, which we obtain from both the superconformal tensor calculus approach and dimensional reduction. For the minimal case, with no vector multiplets, it is known that there is a unique four-derivative superinvariant. However, for the case of one vector multiplet, after field redefinitions, we find that there are three independent superinvariants, one of which is a vector superinvariant that does not contain any curvatures and takes the form of a supersymmetrization of $F^4$. Similarly, for the two vector multiplet case, corresponding to the STU model, we find three gravitational superinvariants and two $F^4$-type vector superinvariants. Moreover, we find that these vector superinvariants do not affect the two- and three-charge static BPS black hole solutions. We further consider the rigid limit to $\mathcal N=2$ super-Yang-Mills and use this to conjecture a family of vector superinvariants for five-dimensional $\mathcal N=2$ supergravity coupled to an arbitrary number of vector multiplets.}

\date{\today}

\begin{document}

\maketitle

\section{Introduction}
Higher-derivative superinvariants have numerous applications from black holes to precision holography. For example, higher-derivative corrected black holes provide a window into quantum gravity~\cite{Cassani:2025sim} and can also shed light on the black hole weak gravity conjecture~\cite{Harlow:2022ich} and the swampland. Additionally, higher-derivative corrections in gauged supergravity are instrumental in precision holography~\cite{Bobev:2020egg,Bobev:2021oku,Bobev:2021qxx,Cassani:2022lrk, Cassani:2023vsa,Cassani:2024tvk,Ma:2024ynp}, which allows matching with $1/N$ corrections on the field theory side; for example, they have played an important role in holographic hydrodynamics and bounds on $\eta/s$~\cite{Cremonini:2009sy}, the ratio of the shear viscosity to the entropy density of the plasma.

It is known that, up to field redefinitions, pure five-dimensional $\mathcal N=2$ supergravity has a unique four-derivative superinvariant~\cite{Liu:2022sew,Bobev:2022bjm,Cassani:2022lrk}. It is then natural to ask if this continues to be true if we couple the theory to matter multiplets. However, while pure four-dimensional $\mathcal N=2$ supergravity also enjoys this uniqueness of the four-derivative corrections~\cite{Bobev:2020egg,Bobev:2020zov,Bobev:2021oku}, this is no longer the case for minimal supergravity in six dimensions~\cite{Chang:2022urm}. But, pure minimal supergravity in six dimensions reduces to the STU model, \emph{i.e.}, five-dimensional $\mathcal N=2$ supergravity coupled to two vector multiplets. Hence, we already see that there are at least two superinvariants for the STU model. Thus, our goal is to determine how many independent superinvariants there are for the STU model and its truncation to a single vector multiplet.

One of the main techniques for constructing superinvariants from the bottom up is the superconformal tensor calculus approach, which involves gauging the superconformal algebra. In particular, this method has been used to great effect in constructing higher-derivative invariants in four-dimensional $\mathcal N=2$~\cite{deWit:1979dzm,deWit:1980lyi,deWit:1984rvr,deWit:1984wbb,LopesCardoso:2000qm,deWit:2006gn,Butter:2013lta}, five-dimensional $\mathcal N=2$~\cite{Hanaki:2006pj,Bergshoeff:2011xn,Ozkan:2013nwa,Ozkan:2013uk,Baggio:2014hua,Ozkan:2016csy,Gold:2023ymc,Gold:2023dfe,Gold:2023ykx}, and six-dimensional $\mathcal N=(1,0)$~\cite{Bergshoeff:1986wc,Bergshoeff:2012ax,Butter:2016qkx,Butter:2017jqu,Novak:2017wqc,Butter:2018wss} supergravities. This approach has the advantage of being off-shell, which leaves the supersymmetry transformations undeformed but at the cost that one must integrate out the auxiliary fields to get to the Poincar\'e frame. On general grounds, one expects at least three four-derivative superinvariants corresponding to the supersymmetrization of $R_{\mu\nu\rho\sigma}^2$, $R_{\mu\nu}^2$, and $R^2$, although these will not necessarily all be unique after field redefinitions are taken into account. However, in five dimensions, there are two choices of multiplet that one may use: the standard Weyl and the dilaton Weyl multiplet. So, in principle, each curvature-squared has two potentially different supersymmetrizations. Even then, we still have the freedom to choose which multiplet to use as a compensator in the construction, namely either a linear multiplet or a hypermultiplet. This is by no means exhaustive, but many invariants have been constructed via the superconformal tensor calculus prescription.

From the top down, we may start with a higher-dimensional four-derivative action and dimensionally reduce. One natural candidate is the ten-dimensional four-derivative action for heterotic supergravity~\cite{Bergshoeff:1988nn,Bergshoeff:1989de}, which was reduced on a $d$-dimensional torus in~\cite{Eloy:2020dko,Elgood:2020xwu,Ortin:2020xdm,Jayaprakash:2024xlr}. This generates half-maximal supergravity in $10-d$ dimensions coupled to $d$ vector multiplets. In five dimensions, this was truncated to $\mathcal N=2$ supergravity coupled to two vector multiplets, and, after dualization of the three-form flux, gives higher-derivative corrections to the STU model~\cite{Cai:2025yyv}. Alternatively, the reduction of minimal supergravity in six dimensions automatically yields the STU model in five dimensions. This is also a natural model to consider since, at the level of the algebra, the six-dimensional Weyl multiplet reduces to the five-dimensional one constructed using the dilaton Weyl multiplet, after appropriate truncation~\cite{Kugo:2000hn}. In particular, it has been shown off-shell that, after appropriate truncation, the six-dimensional Riemann-squared invariant reduces to the five-dimensional dilaton Weyl one~\cite{Bergshoeff:2011xn}. Within the six-dimensional minimal theory, it is known that there are two inequivalent four-derivative actions~\cite{Bergshoeff:1986wc,Bergshoeff:2012ax,Butter:2016qkx,Butter:2017jqu,Novak:2017wqc,Butter:2018wss}, of which one is field redefinition equivalent to the heterotic action~\cite{Chang:2022urm} and the other of which will reduce to a new invariant.

Of course, with such an abundance of higher-derivative invariants, a na\"ive counting would suggest eight or more families of superinvariants. However, because we are working with higher-derivative actions, there is naturally an ambiguity due to our freedom to perform field redefinitions. This ambiguity may be fixed by going to the ``minimal'' field redefinition frame, where there are no derivatives of field strengths and at most single derivatives of scalars. Indeed, we find that many of these superinvariants are related to one another, some of which will turn out to be vector invariants that do not involve the $\mathcal N=2$ supergravity multiplet, except for the mandatory minimal coupling to gravity.

Naturally, black holes are gravitational objects, and so we expect that a vector invariant may potentially leave the solution for BPS black holes invariant. One hint in this direction comes from the fact that the entropy of static three-charge BPS black holes in the heterotic theory~\cite{Cai:2025yyv} matches that of the STU model~\cite{Castro:2007hc}, constructed using the standard Weyl multiplet supersymmetrization of Weyl-squared~\cite{Hanaki:2006pj}. Indeed, we find that the vector invariants do not affect the two- and three-charge static BPS black hole solutions.

Finally, since the vector invariants are only minimally coupled to gravity, it is natural to take a rigid limit by freezing out the graviton multiplet to obtain supersymmetric invariants in $\mathcal N=2$ super-Yang-Mills coupled to $n_v$ vector multiplets. One expects this rigid limit to freeze out any gravitational superinvariants while preserving the vector superinvariants, and, as such, can be applied to any four-derivative superinvariant to isolate the vector superinvariant. Schematically,
\begin{equation}
    \mathcal L_{\partial^2}^\mathrm{sugra}+\alpha_1\mathcal L_{\partial^4}^\mathrm{grav}+\alpha_2\mathcal L_{\partial^4}^\mathrm{vec}\ \overset{\text{rigid}}{\longrightarrow}\ \mathcal L_{\partial^2}^\mathrm{SYM}+\alpha_2\mathcal L_{\partial^4}^\mathrm{vec}.
\end{equation}
For the $n_v=1$ case, applying the rigid limit to the Weyl-squared action constructed using the Weyl multiplet recovers the vector superinvariant found previously. However, for $n_v>1$, we find that this leads to a new family of four-derivative vector superinvariants, which we conjecture to be supergravity invariants.

The rest of this paper is organized as follows. In Section~\ref{sec:superinv}, we obtain four-derivative superinvariants from the superconformal tensor calculus and dimensional reduction, and we field redefine them to the ``minimal'' redefinition frame. In Section~\ref{sec:compareInv}, we classify the independent superinvariants into gravitational and vector invariants, and in Section~\ref{sec:bpsBHs}, we show that the vector superinvariants do not affect static BPS black holes. In Section~\ref{sec:rigid}, we take the rigid limit of the Weyl-squared action and use this to conjecture a family of vector superinvariants. Finally, we conclude in Section~\ref{sec:disc2}. Some specifics of the field redefinitions are relegated to Appendices~\ref{app:fieldReds}, \ref{app:dwfield}, and \ref{app:6dim}.

\section{Constructing four-derivative superinvariants}\label{sec:superinv}

Since our focus is on both bottom-up and top-down constructions of four-derivative superinvariants in $D=5$, $\mathcal N=2$ supergravity, we start with a brief overview of the two-derivative theory.  While the $\mathcal N=2$ theory admits couplings to vector multiplets, tensor multiplets, and hypermultiplets, we will not consider any hypermultiplets in the following.  Moreover, as tensors can be dualized to vectors, our main focus will be on ungauged $\mathcal N=2$ supergravity coupled to $n_v$ vector multiplets.  The bosonic fields of this theory consist of the metric $g_{\mu\nu}$, $n_v+1$ vectors $A_\mu^I$, and $n_v$ (unconstrained) scalars $\varphi_i$.  

Two derivative $\mathcal N=2$ supergravity coupled to $n_v$ vector multiplets has been extensively studied, and its scalar sector is described by very special geometry.  An important feature here is the introduction of $n_v+1$ constrained scalars $X^I$, satisfying the constraint $\mathcal V=1$, where
\begin{equation}
    \mathcal V=C_{IJK}X^IX^JX^K,
\end{equation}
and $C_{IJK}$ is a constant symmetric tensor.  The bosonic two-derivative Lagrangian takes the form
\begin{equation}
    e^{-1}\mathcal{L}_{\partial^2}=R-\frac{3}{2} a_{I J} \partial_\mu X^I \partial^\mu X^J-\frac{3}{4} a_{I J} F_{\mu \nu}^I F^{J \mu \nu}+\frac{1}{4} C_{I J K} \epsilon^{\mu \nu \rho \sigma \lambda} F_{\mu \nu}^I F_{\rho \sigma}^J A_\lambda^K,
\label{eq:C2d}
\end{equation}
where
\begin{equation}
    a_{IJ}=-\left.\fft13\fft\partial{\partial X^I}\fft\partial{\partial X^J}\log\mathcal V\right|_{\mathcal V=1}=3X_IX_J-2C_{IJK}X^K,\qquad X_I=C_{IJK}X^JX^K.
\label{eq:aIJdef}
\end{equation}
From these expressions, it is easy to see that
\begin{equation}
    X_IX^I=1,\qquad X_I=a_{IJ}X^J,\qquad \dd X_I=-a_{IJ}\dd X^J,
\end{equation}
along with
\begin{equation}
    X_I\dd X^I=0,\qquad X^I\dd X_I=0.
\end{equation}
A natural place where this theory arises is in the context of M-theory compactified on a Calabi-Yau threefold, in which case the $C_{IJK}$ are identified as the triple intersection numbers on the threefold.

Much of our focus will be on the STU model, which corresponds to $n_v=2$ and $C_{123}=1/6$.  In this case, we find
\begin{equation}
    X_I=\fft1{3X^I},\qquad a_{IJ}=\fft13\mathrm{diag}\left(\fft1{(X^1)^2},\fft1{(X^2)^2},\fft1{(X^3)^2}\right).
\label{eq:STUscalars}
\end{equation}
Since $X^1X^2X^3=1$, we can choose a convenient parametrization of $X^I$ in terms of two unconstrained scalars $\varphi_1$ and $\varphi_2$ according to
\begin{equation}
    X^1=e^{-\fft1{\sqrt6}\varphi_1-\fft1{\sqrt2}\varphi_2},\qquad X^2=e^{-\fft1{\sqrt6}\varphi_1+\fft1{\sqrt2}\varphi_2},\qquad X^3=e^{\fft2{\sqrt6}\varphi_1}.
\label{eq:stuvarphi}
\end{equation}
In this case, the two-derivative STU model Lagrangian can be written as
\begin{align}
    e^{-1}\mathcal L_{\partial^2}&=R-\fft12\partial_\mu\varphi_1^2-\fft12\partial_\mu\varphi_2^2-\fft1{4(X^1)^2}(F_{\mu\nu}^1)^2-\fft1{4(X^2)^2}(F_{\mu\nu}^2)^2-\fft1{4(X^3)^2}(F_{\mu\nu}^3)^2\nn\\
    &\quad+\fft14 \epsilon^{\mu\nu\rho\sigma\lambda}F_{\mu\nu}^1F_{\rho\sigma}^2A_\lambda^3.
\label{eq:STUmodel}
\end{align}
We can further truncate the STU model by setting $X^1=X^2$ (corresponding to $\varphi_2=0$) and $A_\mu^1=A_\mu^2$, followed by the relabeling $\{X^3,A_\mu^3\}\to\{X^2,A_\mu^2\}$.  The truncated Lagrangian then takes the form
\begin{equation}
    e^{-1}\mathcal L_{\partial^2}=R-\fft12\partial_\mu\varphi_1^2-\fft1{2(X^1)^2}(F_{\mu\nu}^1)^2-\fft1{4(X^2)^2}(F_{\mu\nu}^2)^2+\fft14\epsilon^{\mu\nu\rho\sigma\lambda}F_{\mu\nu}^1F_{\rho\sigma}^1A_\lambda^2.
\label{eq:C112model}
\end{equation}
This corresponds in the language of very special geometry to $n_v=1$ and $C_{112}=1/3$.  We will refer to this truncation as the ``$C_{112}$ model.''

In the following, we present a survey of the known four-derivative superinvariants that can be added to the two-derivative theory.  These superinvariants have been constructed using various techniques, including dimensional reduction from string theory and superconformal tensor calculus methods.  While the latter naturally gives rise to off-shell formulations of the superinvariants, since we wish to highlight the physical couplings, we will present everything as on-shell invariants.  In doing so, we will make use of field redefinitions as appropriate in order to have a uniform framework for comparison of the various invariants.

Before proceeding, we first present the list of superinvariants that we will investigate:
\begin{itemize}
    \item The Weyl-squared invariant constructed from the standard Weyl multiplet~\cite{Hanaki:2006pj}.
    \item The Weyl-squared invariant constructed from the dilaton Weyl multiplet~\cite{Ozkan:2013uk}.
    \item The Ricci tensor-squared invariant constructed from the dilaton Weyl multiplet.  This can be obtained by taking the difference of the Riemann-squared and Weyl-squared invariants constructed in~\cite{Ozkan:2013uk}.
    \item The STU model Riemann-squared invariant obtained by reducing heterotic supergravity on a torus~\cite{Cai:2025yyv}.
    \item A four-derivative vector invariant by reducing a six-dimensional $\mathcal N=(1,0)$ tensor invariant.
\end{itemize}
The Weyl-squared and Riemann-squared invariants manifestly involve the gravity multiplet, while the other two can be viewed as vector multiplet invariants.  We now examine these superinvariants in turn, beginning with the standard Weyl multiplet construction.

\subsection{The standard Weyl multiplet}
We first turn our attention to superinvariants constructed using the standard Weyl multiplet in the superconformal tensor calculus formalism. In the construction of superinvariants, the standard Weyl multiplet is coupled to a vector multiplet and either a linear multiplet or a hypermultiplet as a compensator. However, due to the on-shell nature of the hypermultiplet, it is more natural to adopt a linear multiplet as the  compensator~\cite{Ozkan:2024euj}, which will be the case we focus on.%
\footnote{For the Weyl tensor squared invariant obtained in~\cite{Hanaki:2006pj}, a hypermultiplet has been used as the compensator. Although explicit checks have not been done, the action may still be fully consistent with supersymmetry, as the hypermultiplet does not appear in the higher-derivative action, so its field equations remain inert, which are needed to close its undeformed on-shell supersymmetry transformations.}
The independent fields in the standard Weyl multiplet are
\begin{eqnarray}
\{e_\mu{}^a\ , \psi_\mu^i\ , V_{\mu}{}^{ij}\ , T_{ab} \ , \chi^i \ , D\ , b_\mu \} \ .
\label{5DN2WeylMultiplets}
\end{eqnarray}
Here, $e_\mu{}^a$ is the f\"unfbein, $b_\mu$ is the gauge field for dilatation (which will be set to $0$ after fixing the local special conformal symmetry), $V_\mu{}^{ij}$ is the
$SU(2)$ $R$-symmetry gauge field, and $\psi_\mu^i$ is the gravitino, which gauges the $Q$-supersymmetry. A real scalar $D$, an anti-symmetric tensor $T_{ab}$, and a symplectic Majorana spinor $\chi^i$ play the role of auxiliary fields. The off-shell linear multiplet consists of the fields 
$\{L_{ij}\ , E_{\mu}\ , N\ , \varphi^i\}$, in which $E_\mu$
and $N$ are auxiliary fields. Vector multiplets consist of the fields $\{A_\mu^I,X^I,\lambda^{Ii},Y_{ij}^I\}$. The off-shell vector multiplet is completed by adding the SU(2) triplet auxiliary fields $Y^{ij}$ together with the dynamical fields $A_{\mu}^I$, $X^I$, and $\lambda^{I\,i}$.

The supersymmetrization of the Weyl tensor squared was constructed in~\cite{Hanaki:2006pj}, the supersymmetrization of the Ricci tensor squared term was given in~\cite{Gold:2023ykx}, and the supersymmetrization of $R^2$ was constructed in~\cite{Ozkan:2013nwa}.\footnote{These curvature squared invariants formulated in superspace were given in~\cite{Butter:2014xxa}.} Additionally, so-called off-diagonal invariants were constructed in~\cite{Ozkan:2016csy}. However, after appropriate field redefinitions, the Ricci scalar squared and off-diagonal invariants are both zero in the ungauged limit. Thus, we will focus on the Weyl-squared invariant. 
The Weyl-squared action of~\cite{Hanaki:2006pj} is constructed off-shell, so the auxiliary fields must be integrated out in order to use it for physical applications. This was done in Ref.~\cite{Cassani:2024tvk}, which obtained the on-shell action for five-dimensional $\mathcal N=2$ supergravity coupled to $n_v$ vector multiplets.  At the two-derivative level, the bosonic terms in the Lagrangian are presented above in (\ref{eq:C2d}), while at the four-derivative level, Ref.~\cite{Cassani:2024tvk} obtained
\begin{align}
\mathcal{L}^{\mathrm{SW}}_{C^2}= & \,\lambda_M X^M \mathcal{X}_{\mathrm{GB}}+D_{I J} C_{\mu \nu \rho \sigma} F^{I \mu \nu} F^{J \rho \sigma}+E_{I J K L} F_{\mu \nu}^I F^{J \mu \nu} F_{\rho \sigma}^K F^{L \rho \sigma}\nn \\
& +\widetilde{E}_{I J K L} F_{\mu \nu}^I F^{J \nu \rho} F_{\rho \sigma}^K F^{L \sigma \mu}+I_{I J K L} \partial_\mu X^I \partial^\mu X^J \partial_\nu X^K \partial^\nu X^L+H_{I J K L} \partial_\mu X^I \partial^\mu X^J F_{\rho \sigma}^K F^{L \rho \sigma} \nn\\
& +\widetilde{H}_{I J K L} \partial_\mu X^I \partial^\nu X^J F^{K \mu \rho} F_{\nu \rho}^L-6 X_I X_J \lambda_K F^{I \mu \alpha} F^{J \nu}{ }_\alpha \nabla_\nu \nabla_\mu X^K\nn \\
& +\frac{3}{4} \lambda_{[I} X_{J]} X_K \epsilon^{\mu \nu \rho \sigma \lambda} \nabla_\alpha F_{\mu \nu}^I F_{\rho \sigma}^J F_\lambda^{K \alpha}+W_{I J K L} \epsilon^{\mu \nu \rho \sigma \lambda} F_{\mu \nu}^I F_\rho^{J \alpha} F_{\sigma \alpha}^K \partial_\lambda X^L\nn \\
& +\frac{1}{2} \lambda_I \epsilon^{\mu \nu \rho \sigma \lambda} R_{\mu \nu \alpha \beta} R_{\rho \sigma}{ }^{\alpha \beta} A_\lambda^I.
\label{eq: Cassani Lagrangian}
\end{align}
Here, $\mathcal{X}_\mathrm{GB}=R^{\mu\nu\rho\sigma}R_{\mu\nu\rho\sigma}-4R^{\mu\nu}R_{\mu\nu}+R^2$ is the Gauss-Bonnet invariant, $C_{\mu\nu\rho\sigma}$ is the Weyl tensor, $\lambda_M$ is a set of $n_v+1$ constants parametrizing the four-derivative corrections, and the various tensors, $D_{IJ}$, $E_{IJKL}$, $\widetilde E_{IJKL}$, $I_{IJKL}$, $H_{IJKL}$, $\widetilde H_{IJKL}$ and $W_{IJKL}$, are given in Ref.~\cite{Cassani:2024tvk} and presented in (\ref{eq:tensors}) in Appendix~\ref{app:fieldReds} for completeness.  These tensors are built from the scalars $X^I$ and are linear in the four-derivative couplings $\lambda_M$.  In the Calabi-Yau context, these couplings correspond to the second Chern class numbers of the Calabi-Yau.

Note that the four-derivative Lagrangian, (\ref{eq: Cassani Lagrangian}), includes two terms involving second derivatives of the fields, namely one with $\nabla_\nu\nabla_\mu X^K$ and another with $\nabla_\alpha F_{\mu\nu}^I$.  These terms can be removed by field redefinitions, as shown in Appendix~\ref{app:fieldReds}.  In addition, we choose to work in a field redefinition basis where $\mathcal X_{\mathrm{GB}}$ is replaced by $R_{\mu\nu\rho\sigma}^2$ and $C_{\mu\nu\rho\sigma}$ by $R_{\mu\nu\rho\sigma}$ in order to facilitate comparison among invariants.  After this set of field redefinitions, we end up with the four-derivative Lagrangian
\begin{align}
    \mathcal{L}^{\mathrm{SW}}_{C^2}= & \,\lambda_M X^M R_{\mu\nu\rho\sigma}^2 +D_{I J} R_{\mu \nu \rho \sigma} F^{I \mu \nu} F^{J \rho \sigma}+I'_{I J K L} \partial_\mu X^I \partial^\mu X^J \partial_\nu X^K \partial^\nu X^L\nn \\
    &+ E'_{I J K L} F_{\mu \nu}^I F^{J \mu \nu} F_{\rho \sigma}^K F^{L \rho \sigma}  +\widetilde{E}'_{I J K L} F_{\mu \nu}^I F^{J \nu \rho} F_{\rho \sigma}^K F^{L \sigma \mu}+ \nn\\
    & + H'_{I J K L} \partial_\mu X^I \partial^\mu X^J F_{\rho \sigma}^K F^{L \rho \sigma} + \widetilde{H}'_{I J K L} \partial_\mu X^I \partial^\nu X^J F^{K \mu \rho} F_{\nu \rho}^L\nn \\
    & +W'_{I J K L} \epsilon^{\mu \nu \rho \sigma \lambda} F_{\mu \nu}^I F_\rho^{J \alpha} F_{\sigma \alpha}^K \partial_\lambda X^L+\frac{1}{2} \lambda_I \epsilon^{\mu \nu \rho \sigma \lambda} R_{\mu \nu \alpha \beta} R_{\rho \sigma}{ }^{\alpha \beta} A_\lambda^I,
\label{eq:CassaniFinalForm}
\end{align}
where
\begin{align}
D_{I J}= &\, 3 \lambda_I X_J-\frac{9}{2} \lambda_M X^M X_I X_J, \nn\\
E'_{I J K L}= & \fft1{16}\lambda_M X^M\left(11 a_{I J} a_{K L}-12 a_{I K} a_{J L}+18 a_{I J} X_K X_L+36 a_{I K} X_J X_L-81 X_I X_J X_K X_L\right) \nn\\
& +\frac{3}{2}\lambda_Ia_{L(J}X_{K)}+\fft32\lambda_Ma^{MN}C_{NI(J}X_{K)}X_L,\nn \\
\widetilde{E}'_{I J K L}= &\fft18 \lambda_M X^M\left(-24a_{I J} a_{K L}+12 a_{I K} a_{J L}-36 a_{IK}X_J X_L +81 X_I X_J X_K X_L\right)\nn \\
& -6\lambda_Ia_{KL} X_J-\fft32\lambda_Ia_{JL}X_K-\fft32\lambda_Ma^{MN}C_{NIK}X_JX_L, \nn\\
I'_{I J K L}= & \fft34\lambda_M X^M\left(5a_{I J} a_{K L}-4a_{IK} a_{JL}\right), \nn\\
H'_{I J K L}= &\frac{3}{4} \lambda_M X^M\left(3 a_{I J} a_{K L}-8 a_{I K} a_{J L}+3a_{I J} X_K X_L\right)+3\lambda_I a_{JL} X_K-\fft32\lambda_Ka_{IJ}X_L\nn\\
&-\fft32\lambda_Ma^{MN}C_{NIJ}X_KX_L,\nn\\
\widetilde{H}'_{I J K L}= &\, \lambda_M X^M\left(-6a_{I J} a_{K L}+6 a_{I K} a_{J L}+12 a_{I L} a_{J K}\right)-12\lambda_Ia_{J[K}X_{L]}-6\lambda_Ka_{IJ}X_L, \nn\\
W'_{I J K L}= & \fft32\lambda_MX^M\left(4a_{IJ}a_{KL}-3a_{KL}X_IX_J\right)+\fft32\lambda_Ia_{JL}X_K-\frac92 \lambda_J a_{I L} X_K-\frac{3}{2} \lambda_J a_{K L}X_I\nn\\
&+3\lambda_La_{IJ}X_K-3\lambda_Ma^{MN}C_{NJL}X_IX_K.
\label{eq:cfftensors}
\end{align}
Here, $a^{IJ}$ is the matrix inverse of $a_{IJ}$ defined in (\ref{eq:aIJdef}).

\subsection{The dilaton Weyl multiplet}
We now consider superinvariants constructed using the dilaton Weyl multiplet in the superconformal tensor calculus formalism. Different from the standard Weyl multiplet, the independent fields in the dilaton Weyl multiplet are $\{e_\mu{}^a\ , \psi_\mu^i\ , V_{\mu}{}^{ij}\ ,C_\mu\ , B_{\mu\nu} \ , \psi^i \ , \sigma\ , b_\mu \}$, where $C_\mu$ and $B_{\mu\nu}$ are respectively the one-form and two-form gauge fields, and $\sigma$ and $\psi^i$ are the dilaton and dilatino, which can be used to fix the local dilatation symmetry without coupling to an extra compensating matter multiplet. Using the dilaton Weyl multiplet, three four-derivative superinvariants have been constructed, namely the supersymmetrizations of $R_{\alpha\beta\gamma\delta}^2$~\cite{Ozkan:2013nwa}, $C_{\alpha\beta\gamma\delta}^2+\frac{1}{6}R^2$~\cite{Ozkan:2013uk}, and $R^2$~\cite{Gold:2023ymc}. As with the standard Weyl case, the ungauged Ricci scalar squared invariant is zero after field redefinitions.

\subsubsection{The two-derivative theory}

Since the dilaton Weyl multiplet construction is distinct from that of the standard Weyl multiplet, it is instructive to first review the two-derivative action before turning to the four-derivative couplings.  The off-shell dilaton Weyl two-derivative action for supergravity coupled to one tensor multiplet and $n_v$ vector multiplets in the ungauged limit is given by~\cite{Ozkan:2013uk,Ozkan:2013nwa}
\begin{align}
e^{-1} \mathcal{L}_{\partial^2}^{\mathrm{DW}}&= L\left(R-\frac{1}{2} G_{\alpha\beta}^2-\frac{1}{3} h_{\alpha\beta\gamma}^2+2 V_\alpha^{\prime i j} V_{i j}^{\prime \alpha}\right)+L^{-1} \partial_\alpha L \partial^\alpha L-2 L^{-1} E_\alpha E^\alpha \nn\\
&\quad -2 \sqrt{2} E_\alpha V^\alpha-2 N^2 L^{-1}\nn\\
&\quad +\mathfrak a_{I J}\left[Y_{i j}^I Y^{J i j}-\frac{1}{2} \partial_\mu \rho^I \partial^\mu \rho^J-\frac{1}{4}\left(F_{\alpha\beta}^I-\rho^I G_{\alpha\beta}\right)\left(F^{\alpha\beta J}-\rho^J G^{\alpha\beta}\right)\right. \nn\\
&\quad \left.-\frac{1}{8} \epsilon^{\alpha\beta\gamma\delta\epsilon}\left(F_{\alpha\beta}^I-\rho^I G_{\alpha\beta}\right)\left(F_{\gamma\delta}^J-\rho^J G_{\gamma\delta}\right) C_\epsilon-\frac{1}{2} \epsilon^{\alpha\beta\gamma\delta\epsilon}\left(F_{\alpha\beta}^I-\rho^I G_{\alpha\beta}\right) b_{\gamma\delta} \partial_\epsilon \rho^J\right],
\label{eq:DW2lag}
\end{align}
where
\begin{equation}
    F^I=\dd A^I,\qquad G=\dd C,\qquad h=\dd b+\fft12C\land G.
\end{equation}
In particular, there are $n_v+1$ vector fields $\{C,A^I\}$, one tensor $b$, and $n_v+1$ scalars $\{L,\rho^I\}$, where the $\rho^I$ are unconstrained scalars.  The vector multiplet couplings are specified by the constant matrix $\mathfrak a_{IJ}$.  Note that the $i,j$ indices are $SU(2)$ R-symmetry indices and, in contrast to the standard Weyl multiplet case, $I,J,K,\ldots$ run from 1 to $n_v$.  This is due to the fact that the standard Weyl multiplet includes a vector, whereas the dilaton Weyl multiplet trades the vector for a two-form.

The remaining fields in (\ref{eq:DW2lag}) are auxiliary fields.  The equations of motion allow us to remove them by setting
\begin{equation}
    Y^I_{ij}=E_\alpha=V_\alpha=V^{\prime ij}_\alpha=N=0,
\end{equation}
leaving the on-shell action
\begin{align}
e^{-1} \mathcal{L}_{\partial^2}^{\mathrm{DW}}&= L\left(R-\frac{1}{2} G_{\alpha\beta}^2-\frac{1}{3} h_{\alpha\beta\gamma}^2\right)+L^{-1} \partial_\alpha L \partial^\alpha L\nn\\
&\quad +\mathfrak a_{I J}\left[-\frac{1}{2} \partial_\mu \rho^I \partial^\mu \rho^J-\frac{1}{4}\left(F_{\alpha\beta}^I-\rho^I G_{\alpha\beta}\right)\left(F^{\alpha\beta J}-\rho^J G^{\alpha\beta}\right)\right. \nn\\
&\quad \left.-\frac{1}{8} \epsilon^{\alpha\beta\gamma\delta\epsilon}\left(F_{\alpha\beta}^I-\rho^I G_{\alpha\beta}\right)\left(F_{\gamma\delta}^J-\rho^J G_{\gamma\delta}\right) C_\epsilon-\frac{1}{2} \epsilon^{\alpha\beta\gamma\delta\epsilon}\left(F_{\alpha\beta}^I-\rho^I G_{\alpha\beta}\right) b_{\gamma\delta} \partial_\epsilon \rho^J\right].
\end{align}
This corresponds to supergravity coupled to one tensor and $n_v$ vector multiplets.  The tensor may be dualized to a vector by adding a Lagrange multiplier term
\begin{equation}
    e^{-1}\Delta\mathcal L=\frac{2}{3}\epsilon^{\alpha\beta\gamma\delta\epsilon}H_{\alpha\beta}\qty(h_{\gamma\delta\epsilon}-\frac{3}{2}C_\gamma G_{\delta\epsilon}),
\end{equation}
and integrating out $h_{\alpha\beta\gamma}$ to get
\begin{align}
    e^{-1}\mathcal L_{\partial^2}^{\mathrm{DW}}&=LR-\frac{1}{4}L^{-1}H^2-\frac{1}{4}\mathfrak d_{IJ}F^I_{\alpha\beta}F^J_{\alpha\beta}+\frac{1}{2}\mathfrak e_I F^I_{\alpha\beta}G^{\alpha\beta}-\frac{1}{4}\mathfrak f\, G^2-\frac{1}{2}\mathfrak a_{IJ}\rho^IF^J_{\alpha\beta}H^{\alpha\beta}\nn\\
    &\quad+\frac{1}{4L}\mathfrak a_{IJ}\rho^I\rho^J G_{\alpha\beta}H^{\alpha\beta}+L^{-1}(\partial L)^2-\frac{1}{2}\mathfrak a_{IJ}\partial_\mu\rho^I\partial^\mu\rho^J\nn\\
    &\quad-\frac{1}{4}\epsilon^{\alpha\beta\gamma\delta\epsilon}\qty(H_{\alpha\beta}G_{\gamma\delta}C_{\epsilon}+\frac{1}{2}F^I_{\alpha\beta}F^J_{\gamma\delta}C_\epsilon),
\label{eq:DW2dact}
\end{align}
where we have defined
\begin{align}
    \mathfrak d_{IJ}&=\mathfrak a_{IJ}+L^{-1}\mathfrak a_{IK}\mathfrak a_{JL}\rho^K\rho^L,\nn\\
    \mathfrak e_I&=\mathfrak a_{IJ}\rho^J\qty(1+\frac{1}{2}L^{-1}\mathfrak a_{KL}\rho^K\rho^L),\nn\\
    \mathfrak f&=L+\mathfrak a_{IJ}\rho^I\rho^J+\frac{1}{4}L^{-1}\mathfrak a_{IJ}\mathfrak a_{KL}\rho^I\rho^J\rho^K\rho^L.
\end{align}
The Lagrangian now contains $n_v+2$ vector fields and corresponds to supergravity coupled to $n_v+1$ vector multiplets.  We may put this into the language of very special geometry and a cubic prepotential by defining an enlarged index $\tilde I=\{-1,0,I\}$ and setting
\begin{equation}
    \mathcal F^{-1}=G,\qquad \mathcal F^0=H,\qquad \mathcal F^{I}=F^I.
\end{equation}
We now rewrite the Chern-Simons term as
\begin{equation}
    e^{-1}\mathcal L^{\mathrm{CP-odd}}_{\partial^2,\mathrm{DW}}=-\frac{1}{4}\epsilon^{\alpha\beta\gamma\delta\epsilon} C_{\tilde I\tilde J\tilde K}\mathcal F^{\tilde I}_{\alpha\beta}\mathcal F^{\tilde J}_{\gamma\delta}\mathcal A^{\tilde I}_\epsilon,
\end{equation}
to identify
\begin{equation}
    C_{-1,-1,0}=\frac{1}{3},\qquad C_{-1,IJ}=\frac{1}{6}\mathfrak a_{IJ},
\end{equation}
with all other components being zero. Properly recasting the entire action in terms of a cubic prepotential would require an appropriate identification of constrained scalars $X^{\tilde I}$. However, we already see that the dilaton Weyl action always comes with a $C_{-1,-1,0}$ term (or equivalently a $C_{112}$ term, if we shifted the indices to start from one), and so can never describe the STU model. Hence, for our purposes in this paper, we will focus on the case without additional vectors from here on.

Setting $n_v=0$ and redefining $L=e^{-2\varphi}$, the two-derivative dilaton Weyl action, (\ref{eq:DW2dact}), takes the simple form
\begin{equation}
    e^{-1}\mathcal L_{\partial^2}^{\mathrm{DW}}=e^{-2\varphi}\qty[R+4(\partial\varphi)^2-\frac{1}{2}G^2-\frac{1}{4}e^{4\varphi}H^2]-\frac{1}{4}\epsilon^{\alpha\beta\gamma\delta\epsilon}H_{\alpha\beta}G_{\gamma\delta}C_{\epsilon}\,.
\label{eq:DWstringFrame}
\end{equation}
This Lagrangian can be put into the framework of very special geometry by transforming to the Einstein frame through the Weyl rescaling
\begin{equation}
    g_{\mu\nu}\to e^{4\varphi/3}g_{\mu\nu}.
\label{eq:DWweylRescale}
\end{equation}
The two-derivative Lagrangian then becomes
\begin{equation}
    e^{-1}\mathcal L_{\partial^2}^\mathrm{E}=R-\frac{4}{3}(\partial\varphi)^2-\frac{1}{2}e^{-4\varphi/3}G^2-\frac{1}{4}e^{8\varphi/3}H^2-\frac{1}{4}\epsilon^{\alpha\beta\gamma\delta\epsilon}H_{\alpha\beta}G_{\gamma\delta}C_{\epsilon}.
\label{eq:DWeinTwoDeriv}
\end{equation}
It is easy to see that this Einstein frame Lagrangian matches the $C_{112}$ Lagrangian, (\ref{eq:C112model}), under the mapping
\begin{equation}
    F^1_{\mu\nu}=G_{\mu\nu},\qquad F^2_{\mu\nu}=-H_{\mu\nu},\qquad \varphi_1=-2\sqrt{\fft23}\varphi.
\label{eq:DWto112}
\end{equation}
We now turn to the four-derivative superinvariants.

\subsubsection{The four-derivative \texorpdfstring{$C_{\alpha\beta\gamma\delta}^2+\frac{1}{6}R^2$}{Weyl-squared plus Ricci scalar-squared} invariant}

We first consider the supersymmetric completion of $C_{\alpha\beta\gamma\delta}^2+\frac{1}{6}R^2$, which is given by~\cite{Ozkan:2013uk,Ozkan:2013nwa} in the case $n_v=0$,
\begin{align}
    e^{-1}\mathcal L_{C^2+\frac{1}{6}R^2}^\mathrm{DW}&=-\frac{1}{4}R_{\alpha\beta\gamma\delta}^2+\frac{1}{3}R_{\alpha\beta}^2-\frac{1}{12}R^2-\frac{1}{8}\epsilon^{\alpha\beta\gamma\delta\epsilon}R_{\alpha\beta\lambda\tau}R_{\gamma\delta}{}^{\lambda\tau}C_\epsilon\nn\\
    &\quad+\frac{1}{3}R_{\alpha\beta\gamma\delta}\qty(G^{\alpha\beta}G^{\gamma\delta}-\frac{3}{4}e^{2\varphi}G^{\alpha\beta}H^{\gamma\delta}-\frac{1}{8}e^{4\varphi}H^{\alpha\beta}H^{\gamma\delta})-\frac{1}{3}e^{2\varphi}R_{\alpha\beta}H^{\alpha\gamma}G^\beta{}_{\gamma}\nn\\
    &\quad+\frac{1}{3}e^{4\varphi}R^{\alpha\beta}H^2_{\alpha\beta}+\frac{1}{12}e^{2\varphi}RG\cdot H-\frac{1}{12}e^{4\varphi}RH^2-\frac{1}{64}e^{8\varphi}\qty(H^2)^2-\frac{1}{32}e^{8\varphi}H^4\nn\\
    &\quad-\frac{1}{12}e^{6\varphi}(G\cdot H)H^2+\frac{5}{24}e^{6\varphi}GHHH+\frac{1}{6}e^{4\varphi}H^2G^2+\frac{1}{24}e^{4\varphi}(G\cdot H)^2+\frac{1}{12}\qty(G^2)^2\nn\\
    &\quad-\frac{1}{3}e^{4\varphi}GGHH-\frac{1}{12}e^{4\varphi}GHGH-\frac{1}{12}e^{2\varphi}G^2G\cdot H-\frac{1}{2}GGGH-\frac{1}{3}\qty(\nabla G)^2\nn\\
    &\quad+\frac{1}{6}\qty(\nabla (e^{2\varphi}H))^2-\frac{1}{2}G^4\nn\\
    &\quad+\epsilon^{\alpha\beta\gamma\delta\epsilon}\left[e^{4\varphi}H_{\alpha\beta}H_{\gamma\delta}\nabla^\omega G_{\epsilon\omega}-\frac{1}{8}e^{2\varphi}H_{\beta\omega}\nabla_\alpha \qty(e^{2\varphi}H_\gamma{}^\omega) G_{\delta\epsilon}-\frac{1}{6}e^{2\varphi}H_{\alpha\beta}\nabla^\omega G_{\gamma\omega}G_{\delta\epsilon}\right.\nn\\
    &\left.\quad\kern4em-\frac{1}{24}\nabla^\omega G_{\alpha\omega}G_{\beta\gamma}G_{\delta\epsilon}\right],
\label{eq:DWc2+16R2orig}
\end{align}
where
\begin{align}
    G\cdot H=G_{\alpha\beta} H^{\alpha\beta},\qquad  ABCD \equiv A^\alpha{}_\beta B^\beta{}_\gamma C^\gamma{}_\delta D^\delta{}_\alpha,\qquad H^4=H^\alpha{}_\beta H^\beta{}_\gamma H^\gamma{}_\delta H^\delta{}_\alpha.
\end{align}

In order to put this Lagrangian into the framework of the $C_{112}$ model, we first remove the derivatives of the field strengths $G$ and $H$ through a combination of integration by parts and field redefinitions.  Next, we note that this four-derivative Lagrangian completes the two-derivative string frame action, (\ref{eq:DWstringFrame}).  To match the $C_{112}$ model, we transform to the Einstein frame using the Weyl rescaling, (\ref{eq:DWweylRescale}).  The details of these transformations are given in Appendix~\ref{app:dwfield}, and the final result is the dilaton-Weyl Riemann-squared invariant
\begin{align}
    e^{-1}\mathcal L_{\mathrm{E},C^2+\frac{1}{6}R^2}^\mathrm{DW}&=-\frac{1}{4}X^1\Bigl[R_{\alpha\beta\gamma\delta}^2-R_{\alpha\beta\gamma\delta}\tilde F^{1\,\alpha\beta}\tilde F^{2\,\gamma\delta}-\frac{1}{2}R_{\alpha\beta\gamma\delta}\tilde F^{2\,\alpha\beta}\tilde F^{2\,\gamma\delta}\nn\\
    &\quad+\frac{23}{36}((\tilde F^1)^2)^2-\frac{1}{6}(\tilde F^1\cdot\tilde F^2)^2+\frac{7}{18}(\tilde F^1)^2(\tilde F^2)^2+\frac{1}{3}(\tilde F^1\cdot\tilde F^2)(\tilde F^2)^2\nn\\
    &\quad+\frac{1}{18}((\tilde F^2)^2)^2
    -\frac{10}{3}(\tilde F^1)^4+\frac{1}{3}\tilde F^1\tilde F^1\tilde F^1\tilde F^2+\frac{1}{3}\tilde F^1\tilde F^2\tilde F^1\tilde F^2-\frac{4}{3}\tilde F^1\tilde F^1\tilde F^2\tilde F^2\nn\\
    &\quad-\frac{2}{3}\tilde F^1\tilde F^2\tilde F^2\tilde F^2-\frac{5}{24}(\tilde F^2)^4
    +\left(3(\tilde F^1)^2+7(\tilde F^1\cdot\tilde F^2)-(\tilde F^2)^2\right)(\partial\log X^1)^2\nn\\
    &\quad+\left(-8(\tilde F^1)^2_{\alpha\beta}-20\tilde F^1_{\alpha\gamma}F^2_{\beta}{}^\gamma+4(\tilde F^2)^2_{\alpha\beta}\right)\partial^\alpha\log X^1\partial^\beta\log X^1
    +3(\partial\log X^1)^4\nn\\
    &\quad+\epsilon^{\mu\nu\rho\sigma\lambda}\left(2\tilde F^1_{\mu\nu}\tilde F^1_{\rho\alpha}\tilde F^2_\sigma{}^\alpha+\tilde F^2_{\mu\nu}\tilde F^1_{\rho\alpha}\tilde F^2_\sigma{}^\alpha\right)\partial_\lambda\log X^1\Bigr]-\frac{1}{8}\epsilon^{\alpha\beta\gamma\delta\epsilon}R_{\alpha\beta\lambda\tau}R_{\gamma\delta}{}^{\lambda\tau}A^1_\epsilon,
\label{DWc2+16R2EinFinal}
\end{align}
where we have defined
\begin{equation}
    \tilde F^1\equiv e^{\fft1{\sqrt6}\varphi_1}F^1=\fft{F^1}{X^1},\qquad\tilde F^2\equiv e^{-\fft2{\sqrt6}\varphi_1}F^2=\fft{F^2}{X^2}.
\end{equation}
Note that the Riemann-squared term has the same form as that of the standard Weyl result, (\ref{eq:CassaniFinalForm}), provided we take $\lambda_1=-1/4$ and $\lambda_2=0$.  However, as we discuss below, this match does not extend to the remaining non-Riemann terms in the four-derivative Lagrangian.

\subsubsection{The four-derivative Ricci-squared invariant}

In addition to the $C^2+\frac{1}{6}R^2$ invariant, we may also consider the Riemann-squared invariant.  However, since these two invariants both involve the Riemann tensor, we find it convenient to work with the residual Ricci-squared invariant obtained by subtracting the two.  This can be obtained as
\begin{equation}
    \mathcal L_{\mathrm{Ric}^2}^\mathrm{DW}=\frac{1}{2}\qty(\mathcal L_{\mathrm{Riem}^2}^\mathrm{DW}-\mathcal L^\mathrm{DW}_{C_{\alpha\beta\gamma\delta}^2+\frac{1}{6}R^2}),
\end{equation}
and reads
\begin{align}
e^{-1} \mathcal{L}^\mathrm{DW}_{\mathrm{Ric}^2}&= -\frac{1}{6} R_{\alpha\beta}^2+\frac{1}{24} R^2+\frac{1}{6} R^{\alpha\beta} G_{\alpha\beta}^2+\frac{1}{3}e^{2\varphi} R H_{\alpha\beta} G^{\alpha\beta}-\frac{4}{3}e^{2\varphi} R_{\alpha\beta} H^{\alpha\gamma} G^\beta{}_\gamma-\frac{1}{3}e^{4\varphi} R H^2 \nn\\
& \quad-\frac{1}{12} \epsilon^{\alpha\beta\gamma\delta\epsilon} C_\alpha V_{\beta\gamma}{}^{i j} V_{\delta\epsilon i j}+\frac{1}{6} V^{\alpha\beta i j} V_{\alpha\beta i j}-2e^{8\varphi}\left(H^2\right)^2+\frac{16}{3}e^{6\varphi} GHHH\nn\\
&\quad-\frac{4}{3}e^{6\varphi} \qty(G\cdot H)\,H^2  +\frac{2}{3}e^{4\varphi} H_{\alpha\beta} H_{\gamma\delta}\left(G^{\alpha\beta} G^{\gamma\delta}-2 G^{\alpha\gamma} G^{\beta\delta}\right)+\frac{2}{3}e^{4\varphi} H^2 G^2\nn\\
&\quad -\frac{4}{3}e^{4\varphi} GGHH-\frac{1}{3}e^{2\varphi} G^2\qty(G\cdot H) +e^{2\varphi}GGGH-\frac{1}{48}\left(G^2\right)^2-\frac{1}{24} G^4\nn\\
&\quad-\frac{1}{6} \nabla_\gamma G^{\alpha \gamma} \nabla^\beta G_{\alpha\beta}+2 \nabla_\alpha \qty(e^{2\varphi}H_{\beta\gamma}) \nabla^{[\alpha}\qty(e^{2\varphi} H^{\beta\gamma]})\nn\\
&\quad+\frac{1}{48} \epsilon^{\alpha\beta\gamma\delta\epsilon} \nabla^\omega G_{\epsilon\omega}\left(4e^{2\varphi} H_{\alpha\beta}-G_{\alpha\beta}\right)\left(4 e^{2\varphi}H_{\gamma\delta}-G_{\gamma\delta}\right).
\end{align}
After integrating out the auxiliary fields, performing appropriate field redefinitions, and noting that the Bianchi identity for $H$ implies
\begin{equation}
    \nabla_\alpha \qty(e^{2\varphi}H_{\beta\gamma})\nabla^{[\alpha}\qty(e^{2\varphi} H^{\beta\gamma]})=\frac{4}{3}e^{4\varphi}\qty[H^2(\partial\varphi)^2-2H^2_{\alpha\beta}\partial^\alpha\varphi\,\partial^\beta\varphi],
\end{equation}
this straightforwardly becomes
\begin{align}
    e^{-1}\mathcal L_{\mathrm{Ric}^2}^\mathrm{DW}&=-\frac{1}{24}G^4+\frac{1}{96}(G^2)^2-\frac{1}{12}e^{4\varphi}GHGH-\frac{1}{6}e^{2\varphi}GGGH+\frac{1}{24}e^{2\varphi}G^2\qty(G\cdot H)\nn\\
    &\quad+\frac{1}{24}e^{4\varphi}(G\cdot H)^2-\frac{1}{6}e^{4\varphi}GGHH+\frac{1}{48}e^{4\varphi}G^2H^2\nn\\
    &\quad-\frac{1}{6}e^{6\varphi}GHHH+\frac{1}{24}e^{6\varphi}(G\cdot H)\,H^2-\frac{1}{24}e^{8\varphi}H^4+\frac{1}{96}e^{8\varphi}(H^2)^2\nn\\
    &\quad+\frac{1}{6}G^2(\partial\varphi)^2+\frac{1}{3}e^{2\varphi}(G\cdot H)(\partial\varphi)^2+\frac{1}{6}e^{4\varphi}H^2(\partial\varphi)^2-\frac{2}{3}G^2_{\alpha\beta}\partial^\alpha\varphi\partial^\beta\varphi\nn\\
    &\quad-\frac{4}{3}e^{2\varphi}G_{\alpha}{}^{\gamma}H_{\beta\gamma}\partial^\alpha\varphi\partial^\beta\varphi-\frac{2}{3}e^{4\varphi}H^2_{\alpha\beta}\partial^\alpha\varphi\partial^\beta\varphi-\frac{2}{3}(\partial\varphi)^4.
\label{eq:DWRic2}
\end{align}
By construction, this is written in the string frame and corrects the two-derivative action \eqref{eq:DWstringFrame}.

Once again, we may put this into the $C_{112}$ model form by Weyl scaling according to \eqref{eq:DWweylRescale} to transform to the Einstein frame.  After applying the mapping, (\ref{eq:DWto112}), the final result is the two-derivative $C_{112}$ Lagrangian, (\ref{eq:C112model}), corrected by
\begin{align}
    e^{-1}\mathcal L_{\mathrm{E,Ric}^2}^\mathrm{DW}&=\fft1{96}X^1\bigg[((\tilde F^1-\tilde F^2)^2)^2-4(\tilde F^1-\tilde F^2)^4
    +4(\tilde F^1-\tilde F^2)^2(\partial\log(X^1/X^2))^2\nn\\
    &\kern4em\ -16(\tilde F^1_\alpha{}^\gamma-\tilde F^2_\alpha{}^\gamma)(\tilde F^1_{\beta\gamma}-\tilde F^2_{\beta\gamma})\partial^\alpha\log (X^1/X^2)\partial^\beta\log(X^1/X^2)\nn\\
    &\kern4em\ -4(\partial\log(X^1/X^2))^4\bigg].
\label{eq:DWRic2ein}
\end{align}
Note that this can be identified as a quartic vector multiplet invariant minimally coupled to gravity.  Thus, by taking appropriate linear combinations, we see that the minimal dilaton Weyl multiplet construction yields two four-derivative invariants, namely the supergravity invariant, (\ref{DWc2+16R2EinFinal}), and the vector invariant, (\ref{eq:DWRic2ein}).

\subsection{Heterotic supergravity}

String theory is a natural source of supersymmetric higher-derivative couplings.  Since we focus on five-dimensional $\mathcal N=2$ theories, it would be natural to consider M-theory on Calabi-Yau or heterotic string theory on $K3\times S^1$.  Alternatively, we can consider the heterotic string on $T^4\times S^1$ along with a further truncation to the $\mathcal N=2$ sector.  In either of the heterotic cases, an appropriate truncation of the $K3$ or $T^4$ sector results in the five-dimensional STU model, where the three gauge fields arise from the ``momentum'' and ``winding'' fields along the $S^1$ along with the five-dimensional dual of the three-form $H$.

With the STU truncation in mind, we start with ten-dimensional heterotic supergravity.  After truncating the heterotic gauge fields, we are left in the bosonic sector with the fields $(\hat g_{MN},\hat B_{MN},\phi)$ and the four-derivative corrected action~\cite{Bergshoeff:1988nn,Bergshoeff:1989de}
\begin{equation}
    e^{-1}\mathcal L=e^{-2\phi}\left[R(\Omega)+4\qty(\partial_M\phi)^2-\fft1{12}\tilde{\hat H}_{MNP}^2+\fft{\alpha'}8 R_{MNPQ}(\Omega_+)^2\right],
\label{eq:10DfourDerivAct}
\end{equation}
where
\begin{equation}
    \tilde{\hat H}=\dd\hat B-\frac{\alpha'}{4}\omega_{3L}(\Omega_+).
\end{equation}
Here, $\omega_{3L}$ is the Lorentz-Chern-Simons form and $\Omega_+$ is the torsionful spin connection
\begin{equation}
    \Omega_\pm=\Omega\pm\fft12\mathcal H,\qquad\mathcal H^{AB}\equiv \tilde{\hat H}_M{}^{AB}\dd x^M.
\end{equation}
Reducing this action on $T^5$ results in five-dimensional $\mathcal N=4$ supergravity coupled to five $\mathcal N=4$ vector multiplets.  This can be further truncated to $\mathcal N=2$ supergravity coupled to two $\mathcal N=2$ multiplets by keeping only the fields pertaining to a single circle.  Alternatively, one can reduce the theory on $K3\times S^1$ to directly obtain the $\mathcal N=2$ STU model.

The general torus reduction of the four-derivative heterotic action was performed in~\cite{Eloy:2020dko,Elgood:2020xwu,Ortin:2020xdm,Jayaprakash:2024xlr}.  Focusing on a single circle, the string frame reduction ansatz takes the form
\begin{align}
    \dd\hat s^2&=g_{\mu\nu}\dd x^\mu \dd x^\nu+e^{2\sigma}(\dd y+A_\mu \dd x^\mu)^2+\dd s^2(T^4),\nn\\
    \hat B&=\fft12b_{\mu\nu}\dd x^\mu\wedge \dd x^\nu+B_{\mu}\dd x^\mu\wedge(\dd y+A_\mu \dd x^\mu),\nn\\
    \phi&=\varphi+\fft12\sigma,
\label{eq:kkansatz}
\end{align}
where $x^\mu$ are five-dimensional coordinates and $y$ is the circle coordinate.  This leads to the three-form flux
\begin{equation}
    \tilde{\hat H}=\tilde h+ G\land(\dd y+A),
\end{equation}
where
\begin{equation}
    \tilde h=h-\frac{\alpha'}{4}\omega_{3L}(\omega_+),\qquad h=\dd b-B\land F,\qquad G=\dd B,\qquad F=\dd A.
\end{equation}
The five-dimensional fields are the metric $g_{\mu\nu}$, two scalars $\varphi$ and $\sigma$, two gauge fields $A$ and $B$, and the two-form potential $b$ with field strength $h$.

The $O(1,1)$ structure of the action can be highlighted by introducing the `left' and `right' gauge field combinations
\begin{equation}
    F^{(\pm)}=e^{\sigma}F\pm e^{-\sigma}G.
\end{equation}
After dualizing the three-form flux and transforming to the Einstein frame, the resulting Lagrangian is given, after appropriate field redefinitions, by~\cite{Cai:2025yyv}
\begin{align}
    \mathcal{L}^\mathrm{het}\,\dd[5]x =&\;\;\sqrt{-g}\left( R - \dfrac{4}{3} (\partial\varphi)^2 - (\partial\sigma)^2 - \dfrac{1}{8}(\tilde F^{(-)\,2} + \tilde F^{(+)\,2}) - \dfrac{1}{4}\tilde H^2 \right)\dd[5]x\nonumber\\
    &+ \dfrac{1}{4} C \wedge \qty(F^{(+)}\wedge F^{(+)} - F^{(-)} \wedge F^{(-)}) + \dfrac{\alpha'}{8} \, \mathcal{L}^\mathrm{het}_{\partial^4}\,\dd[5]x,
\label{eq:finalL2}
\end{align}
where
\begin{align}
    \mathcal L_{\partial^4}^\mathrm{het}\,\dd[5]x&=\,e^{-4\varphi/3}\biggl[R_{\alpha\beta\gamma\delta}^2+\fft12R^{\alpha\beta\gamma\delta}\qty(\tilde H_{\alpha\beta}\tilde H_{\gamma\delta}-\tilde F^{(-)}_{\alpha\beta}\tilde F^{(-)}_{\gamma\delta})-\fft{128}{27}\qty((\partial\varphi)^2)^2-\fft{80}9(\partial\varphi\cdot\partial\sigma)^2\nn\\
    &\kern5em+\fft{16}9(\partial\varphi)^2(\partial\sigma)^2-\fft49\left(5\tilde F^{(+)\,2}_{\mu\nu}+3\tilde F^{(-)\,2}_{\mu\nu}+12\tilde H^2_{\mu\nu}\right)\partial^\mu\varphi\partial^\nu\varphi\nn\\
    &\kern5em
    +\left(\tilde F^{(+)\,2}_{\mu\nu}-\tilde F^{(-)\,2}_{\mu\nu}-4\tilde H^2_{\mu\nu}\right)\partial^\mu\sigma\partial^\nu\sigma+\fft43\tilde F^{(+)}_{\mu\lambda}\tilde F^{(-)}_{\nu\lambda}\qty(\partial^\mu\sigma\partial^\nu\varphi-\partial^\mu\varphi\partial^\nu\sigma)\nn\\
    &\kern5em+\fft29\qty(3\tilde F^{(+)\,2}+\tilde F^{(-)\,^2}+5\tilde H^2)(\partial\varphi)^2+\fft16\qty(-2\tilde F^{(+)\,2}+4\tilde F^{(-)\,^2}+5\tilde H^2)(\partial\sigma)^2\nn\\
    &\kern5em+\fft23\tilde F^{(+)}_{\mu\nu}\tilde F^{(-)}_{\mu\nu}(\partial\varphi\cdot\partial\sigma)+\fft1{288}\qty(\tilde F^{(+)\,2}+\tilde F^{(-)\,2})\qty(\tilde F^{(+)\,2}+13\tilde F^{(-)\,2})\nn\\
    &\kern5em+\fft1{144}\tilde H^2\qty(23\tilde F^{(+)\,2}+17\tilde F^{(-)\,^2}+44\tilde H^2)+\fft1{16}\left((\tilde H_{\mu\nu}\tilde F^{(+)}_{\mu\nu})^2-(\tilde H_{\mu\nu}\tilde F^{(-)}_{\mu\nu})^2\right)\nn\\
    &\kern5em-\fft1{16}\qty(\tilde F^{(+)\,4}+\tilde F^{(-)\,4})-\fft14\tilde F^{(+)}\tilde F^{(+)}\tilde F^{(-)}\tilde F^{(-)}-\fft18\tilde H\tilde F^{(+)}\tilde H\tilde F^{(+)}\nn\\
    &\kern5em-\fft18\tilde H\tilde F^{(-)}\tilde H\tilde F^{(-)}-\tilde H\tilde H\tilde F^{(+)}\tilde F^{(+)}-\fft12\tilde H\tilde H\tilde F^{(-)}\tilde F^{(-)}-\fft{11}8\tilde H^4\nn\\
    &\kern5em-\fft16\epsilon^{\mu\nu\rho\sigma\alpha}\tilde F^{(+)}_{\mu\nu}\tilde F^{(+)}_{\rho\sigma}\tilde H_{\alpha\beta}\partial^\beta\varphi-\fft14\epsilon^{\mu\nu\rho\sigma\alpha}\tilde F^{(+)}_{\mu\nu}\tilde F^{(-)}_{\rho\sigma}\tilde H_{\alpha\beta}\partial^\beta\sigma\nn\\
    &\kern5em+\fft12\epsilon^{\mu\nu\rho\sigma\alpha}\tilde H_{\mu\nu}\tilde F^{(+)}_{\rho\lambda}\tilde F^{(-)}_{\sigma\lambda}\partial_\alpha\sigma\biggr]\dd[5]x+2C\wedge\Tr[R \wedge R].
\label{eq:finalL4}
\end{align}
Here, we have defined $\tilde F^{(\pm)}\equiv e^{-2\varphi/3}F^{(\pm)}$ and $\tilde H\equiv e^{4\varphi/3}H$, where $H=\dd C$ is the dualized three-form.  We have also introduced the notation $F^4\equiv F_\alpha{}^\beta F_\beta{}^\gamma F_\gamma{}^\delta F_\delta{}^\alpha$ and $ABCD\equiv A_\alpha{}^\beta B_\beta{}^\gamma C_\gamma{}^\delta D_\delta{}^\alpha$ for the quartic field strength combinations.

We can translate the heterotic variables to match that of the two-derivative STU Lagrangian, (\ref{eq:STUmodel}), by taking
\begin{align}
    &X^1=e^{2\varphi/3-\sigma},\qquad X^2=e^{2\varphi/3+\sigma},\qquad X^3=e^{-4\varphi/3},\nn\\
    &F^1=F,\kern4.8em F^2=-G,\kern3.9em F^3=H,
\label{eq:Het-STU relations}
\end{align}
which is equivalent in our notation to setting
\begin{equation}
    \tilde F^{(\pm)}=\tilde F^1\mp\tilde F^2,\qquad\tilde H=\tilde F^3.
\end{equation}
In this case, the four-derivative heterotic Lagrangian, \eqref{eq:finalL4}, takes the form
\begin{align}
    \mathcal{L}_{\partial^4}^\mathrm{het}\,\dd[5]x &= X^3\Bigl[R_{\alpha\beta\gamma\delta}^2 +\dfrac{1}{2}R^{\alpha\beta\gamma\delta} \left(\tilde F^3_{\alpha\beta} \tilde F^3_{\gamma\delta}- (\tilde F^1_{\alpha\beta} + \tilde F^2_{\alpha\beta})(\tilde F^1_{\gamma\delta} + \tilde F^2_{\gamma\delta})\right)\nn\\
    &\quad - \dfrac{128}{27} (\partial \varphi)^2 (\partial \varphi)^2 - \dfrac{80}{9} (\partial\varphi \cdot \partial\sigma)^2 + \dfrac{16}{9}  (\partial \varphi)^2 (\partial \sigma)^2 \nn\\
    &\quad +\dfrac{7}{72}\left( (\tilde F^1)^2 (\tilde F^1)^2 +  (\tilde F^2)^2 (\tilde F^2)^2 + 2 (\tilde F^1)^2 (\tilde F^2)^2 \right) + \dfrac{1}{6} \left((\tilde F^1)^2+(\tilde F^2)^2\right) (\tilde F^1\cdot \tilde F^2)\nn\\
    &\quad+ \dfrac{5}{18} \left((\tilde F^1)^2 + (\tilde F^2)^2\right) (\tilde F^3)^2
    - \dfrac{1}{4} (\tilde F^1\cdot \tilde F^3) (\tilde F^2\cdot F^3) - \dfrac{1}{12} (\tilde F^1\cdot \tilde F^2) (\tilde F^3)^2\nn\\
    &\quad+ \dfrac{11}{36} ((\tilde F^3)^2)^2 -\dfrac{3}{8} \left( (\tilde F^1)^4 +  (\tilde F^2)^4  \right) + \dfrac{1}{4} \tilde F^1 \tilde F^2 \tilde F^1 \tilde F^2 -\dfrac{1}{2} \tilde F^1 \tilde F^2 \tilde F^2 \tilde F^1 \nn\\
    &\quad - \dfrac{1}{4} \left(\tilde F^1 \tilde F^3 \tilde F^1 \tilde F^3 + \tilde F^2 \tilde F^3 \tilde F^2 \tilde F^3\right)- \dfrac{3}{2} \left( \tilde F^1 \tilde F^3 \tilde F^3 \tilde F^1 + \tilde F^2 \tilde F^3 \tilde F^3 \tilde F^2\right) \nn\\
    &\quad + \tilde F^1 \tilde F^2 \tilde F^3 \tilde F^3  - \dfrac{11}{8} (\tilde F^3)^4 \nn \\
    &\quad + \dfrac{2}{9} \left(4\left((\tilde F^1)^2 + (\tilde F^2)^2 - (\tilde F^1\cdot\tilde F^2)\right)+5 (\tilde F^3)^2\right)(\partial\varphi)^2\nn \\
    &\quad + \dfrac{1}{6} \left(2\left((\tilde F^1)^2 + (\tilde F^2)^2\right) +12 (\tilde F^1\cdot\tilde F^2)+5(\tilde F^3)^2\right)(\partial\sigma)^2\nn\\
    &\quad + \dfrac{2}{3} \left( (\tilde F^1)^2 - (\tilde F^2)^2 \right) (\partial \varphi \cdot \partial \sigma)\nn\\
    &\quad - \dfrac{16}{9} \left(2\left( (\tilde F^1)^2_{\alpha\beta} + (\tilde F^2)_{\alpha\beta} \right)-  \tilde F^1_{\alpha\gamma} \tilde F^2_{\beta\gamma}+3 (\tilde F^3)^2_{\alpha\beta} \right) (\partial^\alpha\varphi \partial^\beta \varphi)\nn\\
    &\quad - \dfrac{8}{3} (\tilde F^1_{\alpha\gamma} \tilde F^2_{\beta\gamma} - \tilde F^2_{\alpha\gamma} \tilde F^1_{\beta\gamma}) (\partial^\alpha\varphi \partial^\beta \sigma) - 4 \left(\tilde F^1_{\alpha\gamma} \tilde F^2_{\beta\gamma} + (\tilde F^3)^2_{\alpha\beta} \right) (\partial^\alpha\sigma \partial^\beta\sigma)\nn\\
    &\quad - \dfrac{2}{3} \epsilon^{\alpha\beta\gamma\delta\epsilon} \left(\tilde F^1_{\alpha\beta} \tilde F^1_{\gamma\mu} +  \tilde F^2_{\alpha\beta}\tilde F^2_{\gamma\mu} -\tilde F^2_{\alpha\beta} \tilde F^1_{\gamma\mu} - \tilde F^1_{\alpha\beta}\tilde F^2_{\gamma\mu} \right) \tilde F^3_{\delta}{}^\mu\partial_\epsilon\varphi\nn\\
    &\quad - \epsilon^{\alpha\beta\gamma\delta\epsilon} \left(  ( \tilde F^1_{\alpha\beta}\tilde F^1_{\gamma\mu} - \tilde F^2_{\alpha\beta} \tilde F^2_{\gamma\mu}) \tilde F^3_{\delta}{}^\mu + \tilde F^3_{\alpha\beta}\tilde F^1_{\gamma\mu} \tilde F^2_{\delta}{}^\mu  \right) \partial_\epsilon\sigma\Bigr]\dd[5]x+2A^3\wedge\Tr[R\wedge R].
\label{eq:hetred}
\end{align}
Notice that, in the heterotic reduction, the $H$-field (or, equivalently, $F^3$ in the STU language) always enters in even powers in the CP-even sector and odd powers in the CP-odd sector.  This is in contrast with the standard Weyl supergravity Lagrangian, (\ref{eq:CassaniFinalForm}), where the gauge fields $F^I$ have no particular parity restrictions.  This is a strong hint that the heterotic four-derivative corrections are not of the form previously constructed by superconformal tensor calculus methods.

\subsection{Dimensional reduction from six dimensions}

Finally, another source of four-derivative corrections can be obtained by dimensional reduction of six-dimensional $\mathcal N=(1,0)$ supergravity coupled to one tensor multiplet with bosonic fields $(g_{MN}, \hat B_{MN},\phi)$.  From~\cite{Bergshoeff:1986wc,Bergshoeff:2012ax,Butter:2016qkx,Butter:2017jqu,Novak:2017wqc,Butter:2018wss}, it is known that there are two independent four-derivative superinvariants for minimal supergravity in six dimensions, corresponding to the supersymmetrizations of the Riemann tensor squared and the Gauss-Bonnet term.%
\footnote{There are in fact three superinvariants, but the one corresponding to $R^2$ is proportional to the two-derivative equations of motion, as is also the case in five dimensions.}
The symmetric combination of these superinvariants (Riemann-squared plus Gauss-Bonnet) is, after appropriate field redefinitions and dualization of the three-form flux, equivalent to the Bergshoeff-de Roo (BdR) action in six dimensions~\cite{Chang:2022urm}. Conversely, it was shown that heterotic supergravity reduced on $T^4$ has equal coefficients of the Riemann-squared plus Gauss-Bonnet invariants, after appropriate dualization~\cite{Liu:2013dna}. Hence, since we already have the dimensional reduction of the heterotic action, we will not consider this combination further.

The antisymmetric combination of Riemann-squared minus Gauss-Bonnet, after dualization of the three-form flux, takes the form~\cite{Chang:2022urm}
\begin{align}
    e^{-1}\mathcal L^{6\mathrm D}&=e^{-2\phi}\bigg\{R(\Omega)+4(\partial\phi)^2-\frac{1}{12}\hat H^2\nn\\
    &\kern4em+\alpha\bigg[-\frac{1}{12}\hat H^{MNP}\partial_M\qty(\frac{1}{2}\Omega_N^{AB}\epsilon_{PABQRS}\hat H^{QRS}+\partial^Q\phi\frac{1}{6}\epsilon_{NPQRST}\hat H^{RST})\nn\\
    &\kern6em\ -\frac{1}{2}R_{MN}(\Omega)^2+\frac{1}{8}R(\Omega)^2-\frac{1}{8}R_{MNPQ}(\Omega)\hat H^{MNR}\hat H^{PQ}{}_R+\frac{1}{8}R^{MN}(\Omega)\hat H^2_{MN}\nn\\
    &\kern6em\ -\frac{1}{48}R(\Omega)\hat H^2-4R_{MN}(\Omega)\qty(\hat \nabla^M\hat \nabla^N\phi+\partial^M\phi\partial^N\phi)-R(\Omega)(\partial\phi)^2\nn\\
    &\kern6em\ +\frac{3}{2}R(\Omega)\hat \Box\phi-\frac{1}{2}\hat H^2_{MN}\partial^M\phi\partial^N\phi+\frac{1}{3}\hat H^2(\partial\phi)^2{\color{red}\boldsymbol{-}}\frac{1}{12}\hat H^2\hat \Box\phi-8\qty(\hat \nabla_M\hat \nabla_N\phi)^2\nn\\
    &\kern6em\ +\frac{11}{2}(\hat \Box\phi)^2-16\hat \nabla_M\hat \nabla_N\phi\partial^M\phi\partial^N\phi-14(\partial\phi)^2\hat \Box\phi+10(\partial\phi)^4+\frac{1}{24}(\hat \nabla \hat H)^2\nn\\
    &\kern6em\ -\frac{1}{8}\qty(\hat \nabla^M\hat H_{MNP})^2+\frac{1}{6}\hat H^{MNP}\partial^Q\phi\hat \nabla_Q \hat H_{MNP}+\frac{1}{4}\hat H_{MNP}\partial^M\phi\hat \nabla^Q\hat H_Q{}^{NP}\nn\\
    &\kern6em\ +\frac{1}{24}\hat H^4-\frac{1}{32}(\hat H^2_{MN})^2+\frac{1}{1152}(\hat H^2)^2-\frac{1}{36}\epsilon^{MNPQRS}\hat H_{QRS}\left(\frac{1}{8}\hat H^2_{MT}\hat H_{NP}{}^T\right.\nn\\
    &\kern6em\ \qquad\left.-\frac{3}{2}\hat H_{MN}{}^T\qty(\hat R_{PT}+4\hat \nabla_P\hat \nabla_T\phi+4\partial_P\phi\partial_T\phi)\right)\bigg]\bigg\},
\label{eq:sixdimact}
\end{align}
where
\begin{equation}
    \hat H^2_{MN}=\hat H_{MAB}\hat H_N{}^{AB},\qquad \hat H^4=\hat H_{MNR}\hat H_{PQ}{}^R\hat H^{MPS}\hat H^{NQ}{}_S,
\end{equation}
and the red minus sign corrects an error in Eq.~(3.18) of~\cite{Chang:2022urm}. Here, $\phi$ is the six-dimensional dilaton, $\Omega$ the six-dimensional spin connection, and $\hat H=\dd\hat B$ the six-dimensional three-form flux. In addition, $\hat \nabla$ denotes the six-dimensional covariant derivative, and we are using $M,N,\ldots$ to denote six-dimensional curved indices and $A,B,\ldots$ to denote six-dimensional flat indices. Note that, in obtaining this expression, one must use the identity
\begin{equation}
    \epsilon^{MNPQRS}\hat H_{QRS}\hat H_{MAC}\hat H_{NB}{}^C\hat H_P{}^{AB}=\frac{1}{2}\epsilon^{MNPQRS}\hat H_{QRS}\hat H_{MN}{}^Q\hat H^2_{PQ},\label{eq:SezginIdentity}
\end{equation}
which can be derived from the fact that
\begin{equation}
    \epsilon^{[MNPQRS}\hat H_{QRS}\hat H_{M}{}^{A]}{}_C \hat H_{NB}{}^C\hat H_{PA}{}^{B}=0.
\end{equation}
Notice also that the first line of the four-derivative invariant can be absorbed into a field redefinition of $\hat B$.

Note that this expression contains second derivatives on the fields, such as $\hat\nabla\hat\nabla\phi$ and $\hat\nabla\hat H$.  These terms can be eliminated by field redefinitions, which corresponds to using the two-derivative equations of motion, along with integration by parts and Bianchi identities.  The explicit integration by parts transformations are given in Appendix~\ref{app:6dim}.  This then allows us to write the action in the compact form
\begin{align}
    e^{-1}\mathcal L^{6\mathrm{D}}&=e^{-2\phi}\left\{R(\Omega)+4(\partial\phi)^2-\frac{1}{12}\hat H^2\right.\nn\\
    &\qquad\qquad+\alpha\left[-2(\partial\phi)^4+\frac{1}{6}\hat H^2(\partial\phi)^2-\hat H^2_{MN}\partial^M\phi\partial^N\phi+\frac{1}{96}(\hat H^2)^2-\frac{1}{16}(\hat H_{MN}^2)^2\right.\nn\\
    &\qquad\qquad\qquad\ \,\left.\left.+\frac{1}{24}\hat H^4+\frac{1}{6}\epsilon^{MNPQRS}\hat H_{QRS}\qty(\frac{1}{24}\hat H^2_{MT}\hat H_{NP}{}^T+\hat H^T{}_{NP}\partial_M\phi\partial_T\phi)\right]\right\}.
\label{eq:6dAction}
\end{align}

By subtracting the Gauss-Bonnet invariant from the Riemann-squared invariant, we are left with a four-derivative Lagrangian without any invariant Riemann terms.  This suggests that the four-derivative terms in (\ref{eq:6dAction}) can actually be written as a quartic tensor multiplet coupling.  To see this, we note that the $\mathcal N=(1,0)$ field content is that of a gravity multiplet, $(g_{MN},\hat B^{(+)},\Psi_M)$, coupled to a tensor multiplet, $(\hat B^{(-)},\chi,\phi)$.  In particular, we can decompose the three-form field strength into its self-dual and anti-self-dual parts according to $\star \hat H^{(\pm)}=\pm\hat H^{(\pm)}$.  In this case, the Lagrangian simplifies to
\begin{align}
    e^{-1}\mathcal L^{6\mathrm{D}}&=e^{-2\phi}\qty[R(\Omega)+4(\partial\phi)^2-\frac{1}{12}\hat H^2]\nn\\
    &\qquad-\frac{\alpha}{8}e^{-2\phi}\qty[\frac{4}{3}\qty(\hat H^{(-)})^4+16\qty(\hat H^{(-)})^2_{MN}\partial^M\phi\partial^N\phi+16(\partial\phi)^4],
\label{eq:6dactionCorrected}
\end{align}
where the details of the decomposition are given in Appendix~\ref{app:6dim}.  This matches the IIB one-loop action on $K3$ restricted to the NS-NS sector obtained in~\cite{Liu:2019ses}.  Written in this form, we see that it is manifestly a quartic tensor multiplet invariant of $\mathcal N=(1,0)$ supergravity.

It is also interesting to remark that the IIA one-loop effective action on $K3$ can be obtained by dualizing the heterotic tree-level four-derivative action on $T^4$~\cite{Liu:2013dna}, and, as we have just seen, it seems that the other four-derivative invariant that can be constructed in six dimensions is the dualization of the IIB one-loop effective action. So, in this sense, both invariants that can be constructed for minimal six-dimensional supergravity have a stringy origin as Type II supergravity on $K3$.

\subsubsection{Reduction to five dimensions}

As we are interested in five-dimensional superinvariants, we proceed to reduce the six-dimensional $\mathcal N=(1,0)$ tensor invariant \eqref{eq:6dactionCorrected} on a circle using a standard Kaluza-Klein ansatz
\begin{align}
    \dd\hat s^2&=g_{\mu\nu}\dd x^\mu\dd x^\nu+e^{2\sigma}\eta^2,\qquad \eta=\dd z+A_\mu\dd x^\mu,\nn\\
    \hat B&=\frac{1}{2}b_{\mu\nu}\dd x^\mu\land\dd x^\nu+B_\mu\dd x^\mu\land\eta,\nn\\
    \phi&=\varphi+\frac{1}{2}\sigma.
\end{align}
Note that, although the four-derivative couplings in (\ref{eq:6dactionCorrected}) are written in terms of the anti-self dual $\hat H^{(-)}$, the full six-dimensional theory, namely supergravity coupled to one tensor multiplet, is formulated in terms of an unconstrained $\hat H$.  Reduction of $\hat H$ on a circle gives independent two-form and three-form field strengths in five dimensions.  Dualizing the three-form and taking into account the Kaluza-Klein gauge field as well, we see that the five-dimensional theory has three gauge fields, corresponding to the field content of the STU model.

This resulting five-dimensional action takes the form
\begin{align}
    e^{-1}&\mathcal L^{6\to5}=\nn\\
    &e^{-2\varphi}\left\{R+4(\partial\varphi)^2-(\partial\sigma)^2-\frac{1}{12}h^2-\frac{1}{4}e^{2\sigma}F^2-\frac{1}{4}e^{-2\sigma}G^2\right.\nn\\
    &\qquad+\alpha\left[-2(\partial\phi)^4+\frac{1}{2}(h^2+3e^{-2\sigma}G^2)(\partial\phi)^2-(h^2_{\alpha\beta}+2e^{-2\sigma}G^2_{\alpha\beta})\partial^\alpha\phi\partial^\beta\phi -\frac{1}{288}(h^2)^2\right.\nn\\
    &\qquad\qquad-\frac{1}{48}e^{-2\sigma}h^2G^2+\frac{1}{32}e^{-4\sigma}(G^2)^2-\frac{1}{16}(h_{\alpha\beta}^2)^2 -\frac{1}{8}e^{-2\sigma}h^{\alpha\beta\epsilon}h^{\gamma\delta}{}_\epsilon G_{\alpha\beta}G_{\gamma\delta}\nn\\ 
    &\qquad\qquad+\frac{1}{4}e^{-2\sigma}h^{\alpha\beta\epsilon}h^{\gamma\delta}{}_\epsilon G_{\alpha\gamma}G_{\beta\delta}-\frac{1}{4}e^{-2\sigma}h^2_{\alpha\beta}(G^2)^{\alpha\beta}+\frac{1}{24}h^4-\frac{1}{8}e^{-4\sigma}G^4\nn\\
    &\qquad\qquad+e^{-\sigma}\epsilon^{\mu\nu\rho\sigma\lambda}\left(\frac{1}{48}h^2_{\mu\gamma}h_{\nu\rho}{}^\gamma G_{\sigma\lambda}+\frac{1}{24}e^{-2\sigma}h_{\mu\nu}{}^{\gamma}G^2_{\rho\gamma}G_{\sigma\lambda}+\frac{1}{48}e^{-2\sigma}h_{\mu\alpha\beta}G^{\alpha\beta}G_{\nu\rho}G_{\sigma\lambda}\right.\nn\\
    &\qquad\qquad\qquad\qquad+\frac{1}{72}h_{\rho\sigma\lambda}h^2_{\mu\delta}G_{\nu\delta}+\frac{1}{36}e^{-2\sigma}h_{\rho\sigma\lambda}G^2_{\mu\delta}G_{\nu\delta}-\frac{1}{144}h_{\rho\sigma\lambda}h_{\mu\nu}{}^\gamma h_\gamma{}^{\alpha\beta}G_{\alpha\beta}\nn\\
    &\qquad\qquad\qquad\qquad\left.\left.\left.-\frac{1}{144}e^{-2\sigma}h_{\rho\sigma\lambda}G_{\mu\nu}G^2-\frac{1}{3}h_{\rho\sigma\lambda}G^\tau{}_\nu\partial_\mu\phi\partial_\tau\phi+\frac{1}{2}G_{\mu\nu}h_{\rho\sigma}{}^\tau\partial_\lambda\phi\partial_\tau\phi \right)\right]\right\},
\label{eq:5dAction}
\end{align}
where locally, the three-form and two-form field strengths are
\begin{align}
    h=\dd b-F\land B,\qquad F=\dd A,\qquad G=\dd B.
\end{align}
Note also that
\begin{equation}
    (\partial\phi)^2=(\partial\varphi)^2+\partial\varphi\cdot\partial\sigma+\frac{1}{4}(\partial\sigma)^2.
\end{equation}

One observation is that upon reduction on a circle, one gets $G$'s but not $F$'s in the quartic action,%
\footnote{The exception is the $F\land B$ term in $h$. However, this combination itself is an $O(1,1)$ invariant, and so cannot fix the imbalance between $G$'s and $F$'s.}
meaning that this lacks $O(1,1)$ (or, more generally, $O(d,d)$) invariance. This is unlikely to change with field redefinitions since we are already in the minimal%
\footnote{``Minimal'' in the sense that there are no derivatives of field strengths (including the scalar ``field strength'' $\dd\phi$).}
field redefinition frame, and there is little field redefinition freedom left, especially since all of the two-derivative equations of motion involve derivatives of the field strengths. This is not at all surprising since this is equivalent to 1-loop IIA on $K3\times S^1$, for which the tree-level T-duality rules no longer apply.

\subsubsection{Dualizing the action}

In order to put the above action in the STU framework, we dualize the three-form $h$ by adding a Lagrange multiplier term
\begin{equation}
    \Delta\mathcal L=C\land(\dd h+F\land G).
\end{equation}
Integrating out $C$ yields the Bianchi identity
\begin{equation}
    \dd h=-F\land G,
\end{equation}
whereas integrating out $h$ gives the dualization relation
\begin{equation}
    h^{\mu\nu\rho} = \dfrac{1}{2} \,  e^{2\varphi}\,\epsilon^{\mu\nu\rho\sigma\lambda}\, H_{\sigma\lambda} + \alpha' \dfrac{\delta \mathcal{L}_{\partial^4}}{\delta h_{\mu\nu\rho}},
\end{equation}
where $H=\dd C$. In the two-derivative action, this leads to
\begin{equation}
    -\dfrac{1}{12} e^{-2\varphi} \hat{h}^2 + \dfrac{1}{12}  \,  e^{2\varphi}\,\epsilon^{\mu\nu\rho\sigma\lambda}\, H_{\mu\nu} \, h_{\rho\sigma\lambda} = -\dfrac{1}{4}\, e^{2\varphi}\, H^2.
\end{equation}
Thus, after dualization, we have the action
\begin{align}
    e^{-1}\mathcal L^{6\to 5}&=e^{-2\varphi}\left\{R+4(\partial\varphi)^2-(\partial\sigma)^2-\frac{1}{4}e^{2\sigma}F^2-\frac{1}{4}e^{-2\sigma}G^2-\frac{1}{4}e^{4\varphi}H^2+\frac{1}{4}e^{2\varphi}\epsilon^{\mu\nu\rho\sigma\lambda}F_{\mu\nu}G_{\rho\sigma}C_\lambda\right.\nn\\
    &\qquad\qquad+\alpha\left[-\frac{1}{8}e^{-4\sigma}G^4+\frac{1}{32}e^{-4\sigma}(G^2)^2-\frac{1}{4}e^{4\varphi-2\sigma}GHGH-\frac{1}{2}e^{2\varphi-3\sigma}GGGH\right.\nn\\
    &\qquad\qquad\qquad\quad\!-\frac{1}{8}e^{2\varphi-3\sigma}G^2\qty(G\cdot H)+\frac{1}{8}e^{4\varphi-2\sigma}(G\cdot H)^2-\frac{1}{2}e^{4\varphi-2\sigma}GGHH\nn\\
    &\qquad\qquad\qquad\quad\!+\frac{1}{16}e^{4\varphi-2\sigma}G^2H^2+\frac{1}{2}e^{6\varphi-\sigma}GHHH-\frac{1}{8}e^{6\varphi-\sigma}(G\cdot H)\,H^2\nn\\
    &\qquad\qquad\qquad\quad\!-\frac{1}{8}e^{8\varphi}H^4+\frac{1}{32}e^{8\varphi}(H^2)^2+\frac{1}{2}e^{-2\sigma}G^2(\partial\phi)^2-e^{2\varphi-\sigma}\qty(G\cdot H)(\partial\phi)^2\nn\\
    &\qquad\qquad\qquad\quad\!+\frac{1}{2}e^{4\varphi}H^2(\partial\phi)^2-2e^{-2\sigma}G^2_{\mu\nu}\partial^\mu\phi\partial^\nu\phi+4e^{2\varphi-\sigma}G_{\mu}{}^{\rho}H_{\nu\rho}\partial^\mu\phi\partial^\nu\phi\nn\\
    &\qquad\qquad\qquad\quad\!-2e^{4\varphi}H^2_{\mu\nu}\partial^\mu\phi\partial^\nu\phi-2(\partial\phi)^4\bigg]\bigg\}.\label{eq:TensorInvariantFull}
\end{align}
Note that, since all the CP-odd terms had an odd number of $h$'s, they all become CP-even after dualization.

However, there is more structure to be seen. In particular, note that Hodge duality in six dimensions exchanges $h$ with $G$,
\begin{equation}
    \hat H=h+G\land\eta\implies \star_6\hat H=e^\sigma(\star_5 h)\land\eta+e^{-\sigma}\star_5 G.
\end{equation}
Thus, we may write
\begin{equation}
    \hat H^{(-)}=\frac{1}{2}(\hat H-\star_6\hat H)=\frac{1}{2}\qty[(h-e^{-\sigma}\star_5 G)+(G-e^\sigma\star_5 h)\land\eta].
\end{equation}
This becomes even simpler after dualizing $h=e^{2\varphi}\star_5 H$,
\begin{equation}
    \hat H^{(-)}=\frac{1}{2}\qty[\star_5(e^{2\varphi}H-e^{-\sigma}G)-(e^{2\varphi}H-e^{-\sigma}G)\land e^\sigma\eta].
\end{equation}
Note that $e^{\sigma}\eta$ is the vielbein $E^{6}$. Thus, we see our invariant may be written much more succinctly as
\begin{align}
    e^{-1}\mathcal L^{6\to 5}&=e^{-2\varphi}\left\{R+4(\partial\varphi)^2-(\partial\sigma)^2-\frac{1}{4}e^{2\sigma}F^2-\frac{1}{4}e^{-2\sigma}G^2-\frac{1}{4}e^{4\varphi}H^2+\frac{1}{4}e^{2\varphi}\epsilon^{\mu\nu\rho\sigma\lambda}F_{\mu\nu}G_{\rho\sigma}C_\lambda\right.\nn\\
    &\qquad\qquad+\alpha\left[-\frac{1}{8}(\bar H^{(-)})^4+\frac{1}{32}\qty((\bar H^{(-)})^2)^2+\frac{1}{2}e^{-2\sigma}(\bar H^{(-)})^2(\partial\phi)^2-2(\bar H^{(-)})^2_{\mu\nu}\partial^\mu\phi\partial^\nu\phi\right.\nn\\
    &\qquad\qquad\qquad\,-2(\partial\phi)^4\bigg]\bigg\},\label{eq:niceFormTinv}
\end{align}
where
\begin{equation}
    \bar H^{(-)}_{\mu\nu}=e^{2\varphi}H_{\mu\nu}-e^{-\sigma}G_{\mu\nu}.
\end{equation}
Finally, we may transform to the Einstein frame by doing a Weyl scaling
\begin{equation}
    g_{\mu\nu}\to e^{4\varphi/3}g_{\mu\nu},
\end{equation}
which leaves us with the action
\begin{align}
    e^{-1}\mathcal L_\mathrm{E}^{6\to 5}&=R-\frac{4}{3}(\partial\varphi)^2-(\partial\sigma)^2-\frac{1}{4}(\tilde F^1)^2-\frac{1}{4}(\tilde F^2)^2-\frac{1}{4}(\tilde F^3)^2+\frac{1}{4}\epsilon^{\mu\nu\rho\sigma\lambda}F^1_{\mu\nu}F^2_{\rho\sigma}A^3_\lambda\nn\\
    &\quad+\fft\alpha{32}X^3\left[((\tilde H^{(-)})^2)^2-4(\tilde H^{(-)})^4+4(\tilde H^{(-)})^2(\partial\log(X^3/X^2))^2\right.\nn\\
    &\kern5em-16(\tilde H^{(-)})^2_{\mu\nu}\partial^\mu\log(X^3/X^2)\partial^\nu\log(X^3/X^2)-4(\partial\log(X^3/X^2))^4\bigg],
\label{eq:niceFormTinvein}
\end{align}
where
\begin{align}
    &X^1=e^{2\varphi/3-\sigma},\qquad X^2=e^{2\varphi/3+\sigma},\qquad X^3=e^{-4\varphi/3},\nn\\
    &F^1=F,\kern4.8em F^2=G,\kern4.6em F^3=H,
\label{eq:Het-STU relations 1}
\end{align}
and
\begin{equation}
    \tilde F^I=\fft{F^I}{X^I},\qquad\tilde H^{(-)}=\tilde F^3-\tilde F^2.
\end{equation}
As expected, the reduction of the quartic $\mathcal N=(1,0)$ tensor multiplet coupling gives rise to a vector multiplet invariant in five dimensions.  Furthermore, this invariant has the same structure as that obtained above in (\ref{eq:DWRic2ein}) from the dilaton Weyl multiplet construction.

\subsubsection{IIA vs heterotic frame}

In fact, the connection to the dilaton Weyl multiplet result is not accidental.  To see this, note that the six-dimensional Weyl multiplet is a dilaton Weyl multiplet. Indeed, it is not known how to construct even a two-derivative supergravity model based on the standard Weyl multiplet in six dimensions~\cite{Bergshoeff:1985mz,Coomans:2011ih}. Thus, the six-dimensional Weyl multiplet reduces to the five-dimensional dilaton Weyl multiplet, as long as we set the two vectors that arise in the reduction equal to one another~\cite{Kugo:2000hn}. This is also true at the level of the action, at least for the supersymmetrization of Riemann-squared~\cite{Bergshoeff:2011xn}. We see that if we truncate $\sigma=0$ and $F=\pm G$, then the tensor invariant, \eqref{eq:TensorInvariantFull}, \emph{almost} matches the Ricci-squared invariant constructed using the dilaton Weyl multiplet, \eqref{eq:DWRic2}. This is due to the duality between heterotic supergravity on $T^4$ and IIA supergravity on $K3$. In six dimensions, to get from the IIA frame to the heterotic frame, we must dualize the three-form $H_{(3)}\to\tilde H_{(3)}=\star H_{(3)}$, and then Weyl scale
\begin{equation}
    g_{MN}\to e^{2\phi}g_{MN},
\end{equation}
followed by a flip $\phi\to-\phi$.

In five dimensions, this corresponds to swapping $G_{(2)}\leftrightarrow H_{(2)}$, and then Weyl scaling
\begin{equation}
    g_{\mu\nu}\to e^{2\varphi+\sigma}g_{\mu\nu},
\end{equation}
and, afterwards, rotating the scalar sector
\begin{equation}
    \varphi\to -\frac{1}{2}\varphi-\frac{3}{4}\sigma,\qquad \sigma\to-\varphi+\frac{1}{2}\sigma.
\end{equation}
Note that this latter transformation amounts to $2\varphi+\sigma\to-2\varphi-\sigma$, so overall we map
\begin{equation}
    g^{\mu\nu}\to e^{2\varphi+\sigma}g^{\mu\nu},\qquad\phi\to-\phi,\qquad H^{(-)}_{(2)}\to -e^{-\varphi-\frac{1}{2}\sigma}H^{(-)}_{(2)}.
\end{equation}
We see that performing this dualization leaves the invariant \eqref{eq:niceFormTinv} structurally unchanged, up to a change of the overall scalar factor $e^{-2\varphi}\to e^\sigma$.\footnote{Likewise, applying the scaling transform again returns $e^\sigma\to e^{-2\varphi}$.} Thus, if we transform to the five-dimensional ``IIA frame'' and then truncate $\sigma=0$, $F=\pm G$, we recover the action~\eqref{eq:DWRic2}, as we would expect~\cite{Kugo:2000hn,Bergshoeff:2011xn}.

\section{Four-derivative supergravity and vector invariants}\label{sec:compareInv}

We have focused on four-derivative superinvariants in ungauged $\mathcal N=2$ supergravity coupled to $n_v$ vector multiplets.  As seen in the previous section, these invariants can be obtained through a variety of methods, from superconformal tensor calculus to dimensional reduction of string theory.  The various invariants are summarized in Table~\ref{tbl:invar}.  The standard Weyl, dilaton Weyl, and heterotic invariants involve the supergravity multiplet, while the dilaton Weyl Ricci and 6D invariants are vector multiplet only invariants.  Although we have a total of five invariants, examination of this table suggests that they are not all independent.  To see this in more detail, we may compare the various invariants with each other.

\begin{table}[t]
\begin{tabular}{l|lll}
Invariant&$n_v$&Eqn.&Structure\\
\hline
Weyl&any&(\ref{eq:CassaniFinalForm})&$\lambda_MX^MR_{\mu\nu\rho\sigma}^2+D_{IJ}R^{\mu\nu\rho\sigma}F^I_{\mu\nu}F^J_{\rho\sigma}+\cdots$\\[2pt]
Dilaton Weyl&1&(\ref{DWc2+16R2EinFinal})&$X^1[R_{\mu\nu\rho\sigma}^2-R^{\mu\nu\rho\sigma}(\tilde F^1_{\mu\nu}+\fft12\tilde F^2_{\mu\nu})F^2_{\rho\sigma}+\cdots]$\\[2pt]
DW Ricci&1&(\ref{eq:DWRic2ein})&$X^1[((\tilde F^1-\tilde F^2)^2)^2-4(\tilde F^1-\tilde F^2)^4+\cdots]$\\[2pt]
Heterotic&2&(\ref{eq:hetred})&$X^3[R_{\mu\nu\rho\sigma}^2+\fft12R^{\mu\nu\rho\sigma}(\tilde F^3_{\mu\nu}\tilde F^3_{\rho\sigma}-(\tilde F^1_{\mu\nu}+\tilde F^2_{\mu\nu})(\tilde F^1_{\rho\sigma}+\tilde F^2_{\rho\sigma})+\cdots]$\\[2pt]
6D reduction&2&(\ref{eq:niceFormTinvein})&$X^3[((\tilde F^3-\tilde F^2)^2)^2-4(\tilde F^3-\tilde F^2)^4+\cdots]$
\end{tabular}
\caption{Four-derivative superinvariants in ungauged $\mathcal N=2$ supergravity coupled to $n_v$ vector multiplets.  Note that the dilaton Weyl Ricci construction and the reduction from 6D yield identical vector multiplet invariants up to a relabeling of the fields.}
\label{tbl:invar}
\end{table}

Since the different constructions may involve different numbers of vector multiplets, they cannot necessarily be directly compared.  However, in cases where we can truncate vector multiplets, we can then compare a reduced set of invariants.  Looking at Table~\ref{tbl:invar}, this suggests that we start with pure supergravity, followed by theories with one and two vector multiplets.

\subsection{Pure \texorpdfstring{$\mathcal N=2$}{N=2} supergravity (\texorpdfstring{$n_v=0$}{nv=0})}

For the standard Weyl multiplet construction, we obtain pure supergravity by setting $n_v=0$ along with
\begin{equation}
    C_{111}=1,\qquad X^1=1,\qquad X_1=1,\qquad a_{11}=1.
\end{equation}
In this case, the tensors, (\ref{eq:cfftensors}), reduce to $D_{11}=-3/2$, $E'_{1111}=5/4$ and $\tilde E'_{1111}=-39/8$.  The resulting four-derivative supergravity invariant, (\ref{eq:CassaniFinalForm}), is then
\begin{equation}
    e^{-1}\mathcal L_{\partial^4}=R_{\mu\nu\rho\sigma}^2-\fft32R_{\mu\nu\rho\sigma}F^{\mu\nu}F^{\rho\sigma}+\fft54(F_{\mu\nu}^2)^2-\fft{39}8F^4+\fft12\epsilon^{\mu\nu\rho\sigma\lambda}R_{\mu\nu\alpha\beta}R_{\rho\sigma}{}^{\alpha\beta}A_\lambda.
\label{eq:sugrainv}
\end{equation}
This has a more streamlined presentation in the original field redefinition framework involving the Gauss-Bonnet invariant and the Weyl tensor.  In particular, we have
\begin{align}
    e^{-1}\mathcal L&=R-\fft34F_{\mu\nu}^2+\fft14\epsilon^{\mu\nu\rho\sigma\lambda}F_{\mu\nu}F_{\rho\sigma}A_\lambda\nn\\
    &\quad+
    \mathcal{X}_{\mathrm{GB}}-\fft32C_{\mu\nu\rho\sigma}F^{\mu\nu}F^{\rho\sigma}+\fft98F^4+\fft12\epsilon^{\mu\nu\rho\sigma\lambda}R_{\mu\nu\alpha\beta}R_{\rho\sigma}{}^{\alpha\beta}A_\lambda,
\label{eq:partial4sugra}
\end{align}
where we included the two-derivative terms, (\ref{eq:C2d}), to highlight the normalization of the graviphoton.  In the absence of any vector multiplets, this is the unique four-derivative invariant in minimal $\mathcal N=2$ supergravity~\cite{Liu:2022sew,Bobev:2022bjm,Cassani:2022lrk}.

As can be seen from Table~\ref{tbl:invar}, the only other couplings involving the gravity sector are the dilaton Weyl and heterotic invariants.  However, as these invariants naturally involve a scalar tadpole, $X^1R_{\mu\nu\rho\sigma}^2$ or $X^3R_{\mu\nu\rho\sigma}^2$, it is not possible to decouple the corresponding scalar, which resides in a vector multiplet.  As a result, neither one of these invariants can be truncated to the pure gravity sector~\cite{Liu:2023fqq}.

\subsection{The \texorpdfstring{$C_{112}$}{C112} model (\texorpdfstring{$n_v=1$}{nv=1})}

While very special geometry with one vector multiplet admits several possibilities, we focus on the $C_{112}$ model, (\ref{eq:C112model}), defined by
\begin{equation}
    C_{112}=\fft13\qquad\Rightarrow\qquad (X^1)^2X^2=1,\qquad X_1=\fft2{3X^1},\qquad X_2=\fft1{3X^2}.
\end{equation}
The $a_{IJ}$ matrix is
\begin{equation}
    a_{IJ}=\begin{pmatrix}\fft2{3(X^1)^2}&0\\0&\fft1{3(X^2)^2}\end{pmatrix},
\end{equation}
and the canonically normalized unconstrained scalar in (\ref{eq:C112model}), $\varphi_1$, is given by
\begin{equation}
    X^1=e^{-\fft1{\sqrt6}\varphi_1},\qquad X^2=e^{\fft2{\sqrt6}\varphi_1}.
\end{equation}
We now examine the Riemann-squared invariants of the $C_{112}$ model, starting with the standard Weyl multiplet construction.

\subsubsection{The standard Weyl \texorpdfstring{$C_{112}$}{C112} invariants}

To obtain the standard Weyl four-derivative couplings, we first compute the tensors (\ref{eq:cfftensors}) for $C_{112}=1/3$ and then insert them into the Lagrangian (\ref{eq:CassaniFinalForm}).  The result is a combination of two invariants
\begin{equation}
    \mathcal L_{112}=\lambda_1\mathcal L_{112}^1+\lambda_2\mathcal L_{112}^2,
\end{equation}
parametrized by $\lambda_1$ and $\lambda_2$, where
\begin{align}
    e^{-1}\mathcal L_{112}^1=X^1\biggl[&R_{\alpha\beta\gamma\delta}^2+ \dfrac{1}{2} R_{\alpha\beta\gamma\delta} (-\tilde F^2_{\alpha\beta}\tilde F^2_{\gamma\delta}-2 \tilde F^1_{\alpha\beta}\tilde F^2_{\gamma\delta}) +3((\partial_\alpha\log X^1)^2 )^2\nn\\
    &+ \dfrac{23}{36} ((\tilde F^1)^2)^2+ \dfrac{7}{18} (\tilde F^1)^2 (\tilde F^2)^2 - \dfrac{1}{6} (\tilde F^1 \tilde F^2)^2\nn\\
    &+\dfrac{1}{3} (\tilde F^2)^2 (\tilde F^2 \tilde F^1) + \dfrac{1}{18} ((\tilde F^2)^2)^2
    - \dfrac{10}{3} (\tilde F^1)^4 + \dfrac{1}{3} \tilde F^1 \tilde F^1\tilde F^1\tilde F^2\nn\\
    &+ \dfrac{1}{3} \tilde F^1 \tilde F^2 \tilde F^1 \tilde F^2 - \dfrac{4}{3} \tilde F^1 \tilde F^1 \tilde F^2 \tilde F^2 - \dfrac{2}{3} \tilde F^2 \tilde F^2 \tilde F^2 \tilde F^1- \dfrac{5}{24} (\tilde F^2 )^4\nn\\
    & +\left(3 (\tilde F^1)^2 + 7 (\tilde F^1 \tilde F^2) -  (\tilde F^2)^2\right) (\partial_\alpha\log X^1)^2\nn\\
    &+4(-2\tilde F^1_{\alpha\gamma} \tilde F^1_{\beta\gamma} -5\tilde F^1_{\alpha\gamma} \tilde F^2_{\beta\gamma} +\tilde F^2_{\alpha\gamma} \tilde F^2_{\beta\gamma} )(\partial_\alpha\log X^1)(\partial_\beta\log X^1)\nn\\
    &+\epsilon^{\alpha\beta\gamma\delta\lambda}(2\tilde F^1_{\alpha\beta}+\tilde F^2_{\alpha\beta})\tilde F^1_{\gamma\sigma}\tilde F^2_\delta{}^\sigma\partial_\lambda\log X^1\biggr]
    +\fft12\epsilon^{\alpha\beta\gamma\delta\lambda}R_{\alpha\beta\rho\sigma}R_{\gamma\delta}{}^{\rho\sigma}A_\lambda^1,
\label{eq:SW112l1}
\end{align}
and
\begin{align}
    e^{-1}\mathcal L_{112}^2=X^2\biggl[&R_{\alpha\beta\gamma\delta}^2+ \dfrac{1}{2} R_{\alpha\beta\gamma\delta} (\tilde F^2_{\alpha\beta}\tilde F^2_{\gamma\delta}-4 \tilde F^1_{\alpha\beta}\tilde F^1_{\gamma\delta}) +3((\partial_\alpha\log X^1)^2 )^2\nn\\
    &+ \dfrac{23}{36} ((\tilde F^1)^2)^2 + \dfrac{1}{3} (\tilde F^1)^2(\tilde F^1 \tilde F^2) + \dfrac{11}{36} (\tilde F^1)^2 (\tilde F^2)^2 - \dfrac{7}{12} (\tilde F^1 \tilde F^2)^2\nn\\
    &+\dfrac{1}{3} (\tilde F^2)^2 (\tilde F^2 \tilde F^1) + \dfrac{2}{9} ((\tilde F^2)^2)^2
    - \dfrac{2}{3} (\tilde F^1)^4 - \dfrac{4}{3} \tilde F^1 \tilde F^1\tilde F^1\tilde F^2\nn\\
    &+ \dfrac{1}{6} \tilde F^1 \tilde F^2 \tilde F^1 \tilde F^2 - \dfrac{2}{3} \tilde F^1 \tilde F^1 \tilde F^2 \tilde F^2 - \dfrac{4}{3} \tilde F^2 \tilde F^2 \tilde F^2 \tilde F^1- \dfrac{25}{24} (\tilde F^2 )^4\nn\\
    & +\dfrac12\left(-2 (\tilde F^1)^2 + 12 (\tilde F^1 \tilde F^2) -  (\tilde F^2)^2\right) (\partial_\alpha\log X^1)^2\nn\\
    &-24 \tilde F^1_{\alpha\gamma} \tilde F^2_{\beta\gamma} (\partial_\alpha\log X^1)(\partial_\beta\log X^1)\biggr]
    +\fft12\epsilon^{\alpha\beta\gamma\delta\lambda}R_{\alpha\beta\rho\sigma}R_{\gamma\delta}{}^{\rho\sigma}A_\lambda^2.
\label{eq:SW112l2}
\end{align}
%

\subsubsection{The dilaton Weyl \texorpdfstring{$C_{112}$}{C112} invariant}

The four-derivative dilaton Weyl invariant with $n_v=1$ is given in \eqref{DWc2+16R2EinFinal}.  We see that this is in fact identical to the standard Weyl invariant, (\ref{eq:SW112l1}), with $\lambda_1=-1/4$.  As a result, we conclude that the dilaton Weyl construction does not yield any new four-derivative invariants of the $C_{112}$ model beyond that obtained from the standard Weyl multiplet construction.

\subsubsection{The heterotic \texorpdfstring{$C_{112}$}{C112} truncation}

The final Riemann-squared invariant we consider can be obtained by truncating the heterotic theory.  While the heterotic theory can be written in terms of STU model fields with $n_v=2$, it admits a consistent two-equal-charge truncation by setting $X^1=X^2$ and $F_{\mu\nu}^1=F_{\mu\nu}^2$.  Performing this truncation in the four-derivative Lagrangian, (\ref{eq:hetred}), and relabeling $\{X^3,F_{\mu\nu}^3\}\to\{X^2, F_{\mu\nu}^2\}$ yields the heterotic $C_{112}$ invariant
\begin{align}
    e^{-1}\mathcal{L}_{\mathrm{het}}&= X^2\biggl[R_{\alpha\beta\gamma\delta}^2 +\dfrac{1}{2}R^{\alpha\beta\gamma\delta} \left(\tilde F^2_{\alpha\beta} \tilde F^2_{\gamma\delta}- 4\tilde F^1_{\alpha\beta} \tilde F^1_{\gamma\delta} \right)-24 ((\partial_\alpha\log X^1)^2)^2 \nn\\
    &\kern3em +\dfrac{13}{18} ((\tilde F^1)^2)^2 + \dfrac{17}{36}(\tilde F^1)^2  (\tilde F^2)^2
    - \dfrac{1}{4} (\tilde F^1\cdot \tilde F^2) ^2 + \dfrac{11}{36} ((\tilde F^2)^2)^2 \nn\\
    &\kern3em-(\tilde F^1)^4  - \dfrac{1}{2}\tilde F^1 \tilde F^2 \tilde F^1 \tilde F^2 -2 \tilde F^1 \tilde F^1 \tilde F^2 \tilde F^2  - \dfrac{11}{8} (\tilde F^2)^4 \nn \\
    &\kern3em + \dfrac12 \left(4(\tilde F^1)^2+5 (\tilde F^2)^2\right)(\partial_\alpha\log X^1)^2\nn \\
    &\kern3em -12 \left((\tilde F^1)^2_{\alpha\beta} +(\tilde F^2)^2_{\alpha\beta} \right) (\partial^\alpha\log X^1)(\partial^\beta \log X^1)\biggr]+\fft12\epsilon^{\alpha\beta\gamma\delta\lambda}R_{\alpha\beta\rho\sigma}R_{\gamma\delta}{}^{\rho\sigma}A_\lambda^2.
\end{align}
Comparison with (\ref{eq:SW112l2}) indicates that this heterotic invariant shares the same gravitational terms containing the Riemann tensor.  However, the remaining non-Riemann terms do not match.

We can identify the leftover terms by taking the difference
\begin{align}
    e^{-1}(\mathcal L_{\mathrm{het}}-\mathcal L_{112})&=X^2\biggl[\dfrac{1}{12} ((\tilde F^1-\tilde F^2)^2)^2 -\dfrac13(\tilde F^1-\tilde F^2)^4 
    +3(\tilde F^1-\tilde F^2)^2(\partial_\alpha\log X^1)^2\nn \\
    &\kern3em -12(\tilde F^1_{\alpha\gamma}-\tilde F^2_{\alpha\gamma})(\tilde F^1_{\beta\gamma}-\tilde F^2_{\beta\gamma}) (\partial^\alpha\log X^1)(\partial^\beta \log X^1)\nn\\
    &\kern3em-27 ((\partial_\alpha\log X^1)^2)^2\biggr].    
\end{align}
In addition to the absence of Riemann couplings, this difference depends only on the linear combination $\tilde F^1-\tilde F^2$ that is orthogonal to the graviphoton.  We can additionally make use of the very special geometry constraint $(X^1)^2X^2=1$ to rewrite $\log X^1=\fft13\log(X^1/X^2)$ so that
\begin{align}
    e^{-1}(\mathcal L_{\mathrm{het}}-\mathcal L_{112})&=\fft1{12}X^2\biggl[ ((\tilde F^1-\tilde F^2)^2)^2 -4(\tilde F^1-\tilde F^2)^4 
    +4(\tilde F^1-\tilde F^2)^2(\partial_\alpha\log(X^1/X^2))^2\nn \\
    &\kern3em -16(\tilde F^1_{\alpha\gamma}-\tilde F^2_{\alpha\gamma})(\tilde F^1_{\beta\gamma}-\tilde F^2_{\beta\gamma}) (\partial^\alpha\log(X^1/X^2))(\partial^\beta \log (X^1/X^2))\nn\\
    &\kern3em-4 ((\partial_\alpha\log(X^1/X^2))^2)^2\biggr].
\label{eq:112vec}
\end{align}
It is now clear that this is identical to the vector multiplet invariant, (\ref{eq:DWRic2ein}), obtained from the dilaton Weyl Ricci-squared invariant.  Up to a relabeling of the fields, this also matches the vector multiplet invariant, (\ref{eq:niceFormTinvein}), obtained by dimensional reduction of the six-dimensional $\mathcal N=(1,0)$ tensor invariant.

To summarize, we have explicitly constructed three independent four-derivative invariants of the $C_{112}$ model.  Up to possible linear combinations, these are the two gravitational invariants (\ref{eq:SW112l1}) and (\ref{eq:SW112l2}) along with the vector multiplet invariant (\ref{eq:112vec}).  While there is only a single invariant in the pure supergravity case, the two gravitational invariants here are distinct as they have different mixed couplings to the vector multiplet.  Note that the vector multiplet can be truncated by taking $X^1=X^2=1$ and $F^1=F^2$.  In this case, both gravitational invariants (\ref{eq:SW112l1}) and (\ref{eq:SW112l2}) reduce to the pure supergravity invariant, (\ref{eq:sugrainv}), while the vector multiplet invariant, (\ref{eq:112vec}), vanishes as expected.

\subsection{The STU model (\texorpdfstring{$n_v=2$}{nv=2})}

We now turn to the STU model, which corresponds to $n_v=2$ and $C_{123}=1/6$.  As we have only considered the dilaton Weyl construction with a single vector multiplet, we will restrict our examination to the standard Weyl construction and the heterotic and six-dimensional reduction cases.

\subsubsection{The standard Weyl STU invariant}

The four-derivative STU invariant in the standard Weyl construction can be explicitly obtained by substituting the STU data (\ref{eq:STUscalars}) into the tensors (\ref{eq:cfftensors}) and then writing out the Lagrangian (\ref{eq:CassaniFinalForm}).  The result is lengthy and not particularly illuminating.  Nevertheless, it takes the form
\begin{equation}
    \mathcal L_{\mathrm{STU}}=\lambda_1\mathcal L_{\mathrm{STU}}^1+\lambda_2\mathcal L_{\mathrm{STU}}^2+\lambda_3\mathcal L_{\mathrm{STU}}^3,
\end{equation}
where
\begin{align}
    e^{-1}\mathcal L_{\mathrm{STU}}^3&=X^3\left[R_{\alpha\beta\gamma\delta}^2+\fft12R^{\alpha\beta\gamma\delta}\left(\tilde F^3_{\alpha\beta}\tilde F^3_{\gamma\delta}-(\tilde F^1_{\alpha\beta}+\tilde F^2_{\alpha\beta})(\tilde F^1_{\gamma\delta}+\tilde F^2_{\gamma\delta})\right)+\cdots\right]\nn\\
    &\qquad+\fft12\epsilon^{\alpha\beta\gamma\delta\lambda}R_{\alpha\beta\rho\sigma}R_{\gamma\delta}{}^{\rho\sigma}A_\lambda^3,
\label{eq:LSTU3}
\end{align}
and $\mathcal L_{\mathrm{STU}}^1$ and $\mathcal L_{\mathrm{STU}}^2$ can be obtained by permuting the $I=1,2,3$ indices.

\subsubsection{The heterotic STU invariant}
\label{sec:hetstu}

The Heterotic STU invariant is given in (\ref{eq:hetred}), and is written in a frame where $X^3=e^{-4\varphi/3}$ is directly related to the string dilaton.  Comparison with (\ref{eq:LSTU3}) indicates that all the Riemann terms match.  However, the remaining terms differ.  To see this more explicitly, we can take the difference
\begin{align}
    e^{-1}(\mathcal L_{\mathrm{het}}-\mathcal L_{\mathrm{STU}}^3)&=\fft1{24}X^3\biggl[((\tilde F^1-\tilde F^3)^2)^2-4(\tilde F^1-\tilde F^3)^4+4(\tilde F^1-\tilde F^3)^2(\partial\log(X^1/X^3))^2\nn\\
    &\kern4em-16(\tilde F^1_{\mu\lambda}-\tilde F^3_{\mu\lambda})(\tilde F^1_\nu{}^\lambda-\tilde F^3_\nu{}^\lambda)\partial^\mu\log(X^1/X^3)\partial^\nu\log(X^1/X^3)\nn\\
    &\kern4em-4((\partial\log(X^1/X^3))^2)^2\biggr]\nn\\
    &+\fft1{24}X^3\biggl[((\tilde F^2-\tilde F^3)^2)^2-4(\tilde F^2-\tilde F^3)^4+4(\tilde F^2-\tilde F^3)^2(\partial\log(X^2/X^3))^2\nn\\
    &\kern4em-16(\tilde F^2_{\mu\lambda}-\tilde F^3_{\mu\lambda})(\tilde F^2_\nu{}^\lambda-\tilde F^3_\nu{}^\lambda)\partial^\mu\log(X^2/X^3)\partial^\nu\log(X^2/X^3)\nn\\
    &\kern4em-4((\partial\log(X^2/X^3))^2)^2\biggr].
\label{eq:hetdif}
\end{align}
This clearly demonstrates that the difference is a sum of two vector multiplet invariants.  Note that this reduces to the $C_{112}$ vector invariant, (\ref{eq:112vec}), under the truncation $X^1=X^2$ and $F^1=F^2$.

\subsubsection{Reduction from six dimensions}

Finally, we see that the $\tilde F^2-\tilde F^3$ vector invariant in (\ref{eq:hetdif}) matches the quartic invariant, (\ref{eq:niceFormTinvein}), obtained by dimensional reduction of the $\mathcal N=(1,0)$ tensor couplings.  To summarize, we have identified a set of four-derivative gravity multiplet invariants as well as a set of vector multiplet invariants in the STU model.  The gravity invariants are of the form (\ref{eq:LSTU3}) up to $I=1,2,3$ permutations, while the vector invariants are of the form of the first expression in (\ref{eq:hetdif}), again up to permutations of the indices.

Note that there is some ambiguity in defining the gravity invariants whenever vector multiplets are present, as we can always add an arbitrary combination of vector invariants without changing any of the Riemann couplings.  In particular, the bottom-up Riemann-squared invariant, (\ref{eq:LSTU3}), constructed from superconformal tensor calculus methods, differs from the top-down heterotic Riemann-squared invariant (\ref{eq:hetred}), and the difference arises because of the vector multiplets.  This suggests that higher-derivative observables, such as those for black holes, will depend on the precise manner in which the Riemann-squared invariant is completed.  However, as we now show, this is actually not the case, at least for static BPS black holes.

\section{BPS black holes}\label{sec:bpsBHs}

As we have seen, five-dimensional $\mathcal N=2$ supergravity coupled to $n_v$ vector multiplets admits multiple four-derivative superinvariants.  We now consider the effect of such invariants on black hole solutions.  In particular, we focus on corrections to BPS black holes in the STU model.  The two-derivative STU action,~\eqref{eq:C2d}, admits a well-known static three-charge black hole solution,
\begin{align}
    \dd s^2&=-\frac{\dd t^2}{\mathcal H^{2/3}}+\mathcal H^{1/3}\dd x^i\dd x^i,\qquad\mathcal H=H_1H_2H_3,\nn\\
    A^I&=\frac{1}{H_I}\dd t,\qquad 
    X^I = \dfrac{\mathcal H^{1/3}}{H_I}, \qquad I = 1,2,3.
\label{eq:2der3chargeSol}
\end{align}
This is a multicenter BPS black hole solution written in terms of three harmonic functions satisfying
\begin{equation}
    \partial_i\partial_iH_I=0.
\end{equation}
Additionally, by setting $H_1=H_2$, so that $A^1=A^2$ and $X^1=X^2$, this reduces to a two-charge solution to the $C_{112}$ model \eqref{eq:C112model}.

The four-derivative corrections to the three-charge black hole in heterotic supergravity were obtained in ten dimensions in~\cite{Chimento:2018kop} and reduced to five dimensions in~\cite{Cai:2025yyv}, where the three-form $H$ was dualized into a two-form field strength corresponding to $F^3$ in the STU model language.  The resulting multicenter solution takes the form
\begin{align}
    \dd s^2&=-\fft1{\mathcal H^{2/3}}\qty(1-\alpha'\fft{7(\partial_i \log(H_1/H_2))^2-2\partial_i\log(H_1H_2)\partial_i\log H_3+19(\partial_i\log H_3)^2}{36H_3})\dd t^2\nn\\
    &\qquad+\mathcal H^{1/3}\left(1+\alpha'\fft{(\partial_i \log(H_1/H_2))^2+\partial_i\log(H_1 H_2)\partial_i\log H_3+4(\partial_i\log H_3)^2}{18H_3}\right)\nn\\
    &\kern4em\times\left(\dd x^i\dd x^i-\alpha'\fft{\left(\partial_i\log\mathcal H\dd x^i\right)^2 + 3\partial_{(i}\log\mathcal H\partial_{j)}\log H_3\,\dd x^i\dd x^j}{18H_3}\right),\nn\\
    X^{\hat I} = &\;\; \dfrac{\mathcal H^{1/3}}{H_{\hat I}} \Bigg(1 + \alpha'\; \dfrac{(\partial_i \log\mathcal H)^2}{72 H_3}\Bigg)\qquad(\hat I=1,2), \nn \\
    X^3 = &\;\; \dfrac{\mathcal H^{1/3}}{H_3} \Bigg(1 - \alpha'\; \dfrac{(\partial_i \log\mathcal H)^2}{36 H_3}\Bigg), \nn \\
    A^{\hat I} = &\;\; \dfrac{1}{H_{\hat I}} \Bigg(1 + \alpha' \; \dfrac{\partial_i \log\mathcal H\partial_i\log(H_1H_2)}{12H_3} \Bigg) \dd t\qquad(\hat I=1,2), \nn \\
    A^3 = &\;\;\dfrac{1}{H_3} \Bigg(1 - \alpha' \; \dfrac{\partial_i\log\mathcal H\partial_i\log H_3 }{12H_3}\Bigg)\dd t.
\label{eq:finalSol}
\end{align}
Note that this solution singles out $F^3$ and $X^3$, which correspond to the preferred universal vector multiplet containing the dilaton.

As seen above in Section~\ref{sec:hetstu}, the heterotic invariant differs from the standard Weyl invariant by a set of vector multiplet couplings.  However, we find that the STU model vector invariants, which all have the form \eqref{eq:hetdif}, vanish on-shell for the two-derivative solution.  This is easily seen, for example, by considering the $\tilde F^1-\tilde F^3$ invariant in (\ref{eq:hetdif}).  In particular, for the two-derivative solution, (\ref{eq:2der3chargeSol}), we find explicitly that
\begin{equation}
    \tilde F^1-\tilde F^3=\mathcal H^{-1/3}\dd t\wedge \dd\log(H_1/H_3)=-\mathcal H^{-1/3}\dd t\wedge \dd\log(X^1/X^3).
\end{equation}
This relation also follows from the leading order Killing spinor equations and ensures the vanishing of the four-derivative couplings in (\ref{eq:2der3chargeSol}).  Moreover, the four-derivative equations of motion from the vector invariants vanish.  Hence, these invariants do not lead to any corrections to the leading-order static three-charge BPS solution. 

The vanishing of the vector multiplet corrections immediately implies that the heterotic solution, (\ref{eq:finalSol}), also solves the standard four-derivative STU Weyl action~\eqref{eq:CassaniFinalForm} with ${\lambda_3=\alpha'/8}$.  Using the symmetry between $X_1,X_2,X_3$ in the STU model, it is straightforward to generalize this solution for generic values of $\lambda_I$
\begin{align}
    \dd s^2 &= - \dfrac{1}{\mathcal H^{2/3}} (1 - \delta g_1) \dd t^2 + \mathcal H^{1/3} (1 + \delta g_2) (\dd x^i \dd x^i - \delta g_3),\nn\\
    X^I &= \dfrac{\mathcal H^{1/3}}{H_I} \left(1 + \fft19(\partial_i\log\mathcal H)^2 \sum_{J=1}^3\dfrac{\lambda_J}{H_J} \left(1 - 3 \delta^I_J  \right)\right),\nn\\
    A^I &= \dfrac{1}{H_I} \left(1 + \dfrac{2}{3} \partial_i \log\mathcal H  \sum_{J = 1}^3 \dfrac{\lambda_J}{H_J} ( \partial_i\log(\mathcal H/H_J) - \delta^I_J \partial_i\log\mathcal H\right)\dd t,
\label{eq: generalSTU_BH}
\end{align}
where
\begin{align}
    \delta g_1 &= \fft29\sum_{I=1}^3 \dfrac{\lambda_I}{H_I} \left(7(\partial_i\log(H_J/H_K))^2- 2\partial_i \log(\mathcal H/H_I) \partial_i\log H_I + 19 (\partial_i\log H_I)^2\right),\nn\\
    \delta g_2 &= \fft49\sum_{I=1}^3\dfrac{\lambda_I}{H_I} \left((\partial_i \log(H_J/H_K))^2 + \partial_i\log(\mathcal H/H_I) \partial_i\log H_I + 4 (\partial_i\log H_I)^2\right) ,\nn\\
    \delta g_3 &= \fft49\sum_{I=1}^3\dfrac{\lambda_I}{H_I} \left( (\partial_i\log\mathcal H\,\dd x^i)^2 + 3 \partial_{(i}\log\mathcal  \partial_{j)}\log H_I\, \dd x^i \dd x^j\right),
\end{align}
and where $(I,J,K)$ is an even permutation of $1,2,3$.

Note that the vector invariants do not contribute to the Wald entropy, which is given as
\begin{equation}
    S_\mathrm{W} = -2\pi\int_{\mathcal H} \varepsilon_{\mu\nu}\varepsilon_{\rho\sigma} \dfrac{\partial \mathcal{L}} {\partial R_{\mu\nu\rho\sigma}}\,\dd \Omega_3,
\label{eq:Wald}
\end{equation}
where $\varepsilon$ is the unit binormal to the horizon, as they do not contain any Riemann terms.  Moreover, since the vector invariants vanish on the two-derivative solution, they do not modify the thermodynamics of the BPS black hole.  As a result, the higher-derivative corrected thermodynamics depend only on the gravity invariants parametrized by $\lambda_I$.  In particular, from \eqref{eq:CassaniFinalForm} we have
\begin{align}
    \dfrac{\partial \mathcal{L}}{\partial R_{\mu\nu\rho\sigma}} &= 2 (g^{\mu\rho} g^{\nu\sigma} - g^{\mu\sigma} g^{\nu\rho}) \nn\\
    &\quad+ 8 \left(\lambda_IX^I R^{\mu\nu\rho\sigma} + \dfrac{1}{2} D_{IJ} F^{I \mu\nu} F^{J\mu\nu} + \dfrac{1}{2} \lambda_I \epsilon^{\mu\nu\alpha\beta\gamma} R_{\alpha\beta}^{\quad \rho\sigma} A^I_\gamma\right),
\end{align}
which is universal for the four-derivative gravity multiplet couplings.  The resulting Wald entropy for the higher-derivative corrected STU black hole takes the form~\cite{Castro:2008ys,DominisPrester:2008ynb,Elgood:2020xwu,Cano:2021nzo,Faedo:2019xii,Cai:2025yyv}
\begin{equation}
    S_W = \dfrac{\pi^2}{2G_N} \sqrt{(Q_1 + 24 \lambda_1)(Q_2 + 24\lambda_2)(Q_3 + 24\lambda_3)},
\label{eq: STU entropy final}
\end{equation}
where we have restored Newton's constant and which is valid to linear order in $\lambda_I$.  This result is independent of the four-derivative vector multiplet invariants, and hence is valid for the invariants constructed from the superconformal tensor calculus and for the heterotic string.  In particular, it agrees with the entropy of the heterotic solution that was computed in Ref.~\cite{Cai:2025yyv},
\begin{equation}
    S_\mathrm{W}=\dfrac{\pi^2}{2G_N}\sqrt{Q_1Q_2 (Q_3 + 3\alpha')},\label{eq:HeteroticEntropy}
\end{equation}
after setting $\lambda_1=\lambda_2=0$ and $\lambda_3=\alpha'/8$.

The Wald entropy for four-derivative corrected static BPS black holes in the standard Weyl construction with generic prepotential was obtained in Ref.~\cite{Castro:2007hc}.  Defining the electric charges as
\begin{equation}
    Q_I=-\frac{1}{4\pi^2}\int_{S^3}\dd[3]x\sqrt{-g}\pdv{\mathcal L}{F_{tr}^I},
\end{equation}
they found the Wald entropy to be given by
\begin{equation}
    S_W=\frac{\pi^2}{2G_N}R_H\qty(R_H^2+8\alpha' \lambda_IX^I),\label{eq:STUentropycastro}
\end{equation}
where $X^I$ is evaluated on the horizon and $R_H$ is the horizon radius.
Specializing to the STU model with $C_{123}=1/6$, we can extract the horizon radius from the near-horizon limit of the solution~\eqref{eq: generalSTU_BH}
\begin{equation}
    R_H=(Q_1Q_2Q_3)^{1/6}\left[1+\frac{16\alpha'}{3} \left(\dfrac{\lambda_1}{Q_1}+ \dfrac{\lambda_2}{Q_2}+ \dfrac{\lambda_3}{Q_3}\right)\right].
\end{equation}
Also, from the solution, we find that
\begin{equation}
    X^I\big\vert_{\mathcal H}\overset{\alpha'\to 0}{=} \dfrac{(Q_1Q_2Q_3)^{1/3}}{Q_I} \sim\frac{R_H^2}{Q_I}.
\end{equation}
We then see that the general expression, \eqref{eq:STUentropycastro}, indeed agrees with~\eqref{eq: STU entropy final}, provided $\alpha'$ is absorbed in the couplings $\lambda_I$.

\section{Rigid limit}\label{sec:rigid}

As we have seen, five-dimensional $\mathcal N=2$ supergravity coupled to $n_v$ vector multiplets admits both gravitational and vector invariants at the four-derivative level.  We can have a closer look at the vector invariants by taking the rigid limit of the standard Weyl construction.  Since the natural supergravity invariant, (\ref{eq:partial4sugra}), is given in the field redefinition frame involving $\mathcal X_{\mathrm{GB}}$ and the Weyl tensor, we return to the original form of the standard Weyl multiplet construction with $\mathcal L=\mathcal L_{\partial^2}+\mathcal L_{\partial^4}$, where $\mathcal L_{\partial^2}$ is given in (\ref{eq:C2d}), but with the $\nabla\nabla X^K$ and $\nabla F^I$ terms in (\ref{eq: Cassani Lagrangian}) eliminated by field redefinitions.  In particular, we take
\begin{align}
    e^{-1}\mathcal{L}_{\partial^4}= & \,\lambda_M X^M \mathcal{X}_{\mathrm{GB}} +D_{I J} C_{\mu \nu \rho \sigma} F^{I \mu \nu} F^{J \rho \sigma}+I''_{I J K L} \partial_\mu X^I \partial^\mu X^J \partial_\nu X^K \partial^\nu X^L\nn \\
    &+ E''_{I J K L} F_{\mu \nu}^I F^{J \mu \nu} F_{\rho \sigma}^K F^{L \rho \sigma}  +\widetilde{E}''_{I J K L} F_{\mu \nu}^I F^{J \nu \rho} F_{\rho \sigma}^K F^{L \sigma \mu}+ \nn\\
    & + H''_{I J K L} \partial_\mu X^I \partial^\mu X^J F_{\rho \sigma}^K F^{L \rho \sigma} + \widetilde{H}''_{I J K L} \partial_\mu X^I \partial^\nu X^J F^{K \mu \rho} F_{\nu \rho}^L\nn \\
    & +W'_{I J K L} \epsilon^{\mu \nu \rho \sigma \lambda} F_{\mu \nu}^I F_\rho^{J \alpha} F_{\sigma \alpha}^K \partial_\lambda X^L+\frac{1}{2} \lambda_I \epsilon^{\mu \nu \rho \sigma \lambda} R_{\mu \nu \alpha \beta} R_{\rho \sigma}{ }^{\alpha \beta} A_\lambda^I,
\end{align}
where the scalar tensors are
\begin{align}
D_{I J}= &\, 3 \lambda_I X_J-\frac{9}{2} \lambda_M X^M X_I X_J, \nn\\
E''_{I J K L}= & \fft1{16}\lambda_M X^M\left(-18 a_{I J} a_{K L}-12 a_{I K} a_{J L}+45 a_{I J} X_K X_L+36 a_{I K} X_J X_L-81 X_I X_J X_K X_L\right) \nn\\
& +\frac{3}{4}\lambda_Ia_{LJ}X_{K}-\fft38\lambda_Ia_{LK}X_J+\fft32\lambda_Ma^{MN}C_{NI(J}X_{K)}X_L,\nn \\
\widetilde{E}''_{I J K L}= &\fft18 \lambda_M X^M\left(48a_{I J} a_{K L}+12 a_{I K} a_{J L}-72a_{IJ}X_KX_L-36 a_{IK}X_J X_L +81 X_I X_J X_K X_L\right)\nn \\
& -\fft32\lambda_Ia_{JL}X_K-\fft32\lambda_Ma^{MN}C_{NIK}X_JX_L, \nn\\
I''_{I J K L}= & \fft32\lambda_M X^M\left(a_{I J} a_{K L}+4a_{IK} a_{JL}\right), \nn\\
H''_{I J K L}= &\frac{3}{8} \lambda_M X^M\left(-4 a_{I J} a_{K L}-16 a_{I K} a_{J L}+9a_{I J} X_K X_L\right)+3\lambda_I a_{JL} X_K-\fft94\lambda_Ka_{IJ}X_L\nn\\
&-\fft32\lambda_Ma^{MN}C_{NIJ}X_KX_L,\nn\\
\widetilde{H}''_{I J K L}= &\, \lambda_M X^M\left(12a_{I J} a_{K L}+6 a_{I K} a_{J L}+12 a_{I L} a_{J K}-9a_{IJ}X_KX_L\right)-12\lambda_Ia_{J[K}X_{L]}, \nn\\
W'_{I J K L}= & \fft32\lambda_MX^M\left(4a_{IJ}a_{KL}-3a_{KL}X_IX_J\right)+\fft32\lambda_Ia_{JL}X_K-\frac92 \lambda_J a_{I L} X_K-\frac{3}{2} \lambda_J a_{K L}X_I\nn\\
&+3\lambda_La_{IJ}X_K-3\lambda_Ma^{MN}C_{NJL}X_IX_K .
\end{align}

In order to approach the rigid limit, we take the scalars to have small fluctuations about their vacuum values.  In particular, we let
\begin{equation}
    X^I=\bar X^I+\sqrt2\kappa\varphi^I,\qquad\bar X_I\varphi^I=\mathcal O(\kappa),
\end{equation}
where the constant $\bar X^I$'s satisfy the standard very special geometry identities.  A natural way to decouple the graviphoton at the four-derivative level is to choose moduli
$\bar X_I=\alpha\lambda_I$, where $\alpha$ is a normalization constant that is fixed by the very special geometry constraint $C_{IJK}\bar X^I\bar X^J\bar X^K=1$.  Note that this choice is only possible when all of the $\lambda_I$ couplings have the same sign, and none of them vanish.  Otherwise, we would end up at an unphysical point or on the boundary of moduli space. With this choice, the constant parts of the scalar tensors become
\begin{align}
    \bar D_{IJ}&=-\fft3{2\alpha}\bar X_I\bar X_J,\nn\\
    \bar E''_{IJKL}&=\fft1{16\alpha}\left[-18\bar a_{IJ}\bar a_{KL}-12\bar a_{IK}\bar a_{JL}+33\bar a_{IJ}\bar X_K\bar X_L+42\bar a_{IK}\bar X_J\bar X_L-45\bar X_I\bar X_J\bar X_K\bar X_L\right],\nn\\
    \bar{\tilde E}''_{IJKL}&=\fft1{8\alpha}\left[48\bar a_{IJ}\bar a_{KL}+12\bar a_{IK}\bar a_{JL}-72\bar a_{IJ}\bar X_K\bar X_L-42\bar a_{IK}\bar X_J\bar X_L+63\bar X_I\bar X_J\bar X_K\bar X_L\right],\nn\\
    \bar I''_{IJKL}&=\fft3{2\alpha}\left[\bar a_{IJ}\bar a_{KL}+4\bar a_{IK}\bar a_{JL}\right],\nn\\
    \bar H''_{IJKL}&=\fft3{8\alpha}\left[-4\bar a_{IJ}\bar a_{KL}-16\bar a_{IK}\bar a_{JL}+5\bar a_{IJ}\bar X_K\bar X_L\right],\nn\\
    \bar{\tilde H}''_{IJKL}&=\fft1\alpha\left[12\bar a_{IJ}\bar a_{KL}+6\bar a_{IK}\bar a_{JL}+12\bar a_{IL}\bar a_{JK}-9\bar a_{IJ}\bar X_K\bar X_L\right],\nn\\
    \bar W'_{IJKL}&=\fft3\alpha\left[2\bar a_{IJ}\bar a_{KL}-3\bar a_{KL}\bar X_I\bar X_J\right].
\end{align}

We now decompose the $n_v+1$ gauge fields $A^I$ (with field strengths $F^I$) into the graviphoton and vector multiplet fields
\begin{equation}
    F^I=\bar X^IG+\sqrt2\kappa V^I,\qquad G=\bar X_IF^I,\qquad \sqrt2\kappa V^I=\left(\delta^I_J-\bar X^I\bar X_J\right)F^J.
\end{equation}
In this limit, the two-derivative Lagrangian, (\ref{eq:C2d}), becomes
\begin{align}
    e^{-1}\mathcal L_{\partial^2}&=\fft1{2\kappa^2}\left(R-\fft34G_{\mu\nu}^2+\fft14\epsilon^{\mu\nu\rho\sigma\lambda}G_{\mu\nu}G_{\rho\sigma}\mathcal A_\lambda\right)\nn\\
    &\quad-3\bar a_{IJ}\left(\fft12\partial_\mu\varphi^I\partial^\mu\varphi^J+\fft14V_{\mu\nu}^IV^{J\,\mu\nu}+\fft18\epsilon^{\mu\nu\rho\sigma\lambda}V^I_{\mu\nu}V^J_{\rho\sigma}\mathcal A_\lambda\right)+\mathcal O(\kappa),
\end{align}
where $2\kappa^2$ is the gravitational coupling constant and where $G=\dd\mathcal A$.  At the two-derivative level, the rigid limit is obtained by taking $\kappa\to0$.  After decoupling the gravity sector, we find the simple free matter action
\begin{align}
    \mathcal L&=3\bar a_{IJ}\left(-\fft12\partial_\mu\varphi^I\partial^\mu\varphi^J-\fft14V_{\mu\nu}^IV^{J\,\mu\nu}\right)\nn\\
    &\quad+6\sqrt2\kappa
C_{IJK}\left(\fft12\varphi^I\partial_\mu\varphi^J\partial^\mu\varphi^K+\fft14\varphi^IV_{\mu\nu}^JV^{K\,\mu\nu}+\fft1{24}\epsilon^{\mu\nu\rho\sigma}V_{\mu\nu}^IV_{\rho\sigma}^Ja_\lambda^K\right),
\end{align}
where $V^I=\dd a^I$, and we have restored the $\mathcal O(\kappa)$ couplings.%
\footnote{The classical five-dimensional super-Yang-Mills action typically does not include the cubic $\mathcal O(\kappa)$ terms, but they can be generated in the quantum theory~\cite{Seiberg:1996bd}.}
Although $I$ runs from $1$ to $n_v+1$, the constraints $\bar X_I\partial\varphi^I=0$ and $\bar X_IV^I=0$ ensure that only $n_v$ gauge fields and $n_v$ scalars survive in the rigid limit.

We now turn to the four-derivative Lagrangian, which splits into a sum of three terms
\begin{equation}
    \mathcal L_{\partial^4}=\mathcal L_{G^4}+\mathcal L_{G^2V^2}+\mathcal L_{V^4}.
\end{equation}
The first term is the pure supergravity invariant
\begin{equation}
    2\kappa^2\alpha\,e^{-1}\mathcal L_{G^4}=\mathcal X_{\mathrm{GB}}-\fft32C_{\mu\nu\rho\sigma}G^{\mu\nu}G^{\rho\sigma}+\fft98G^4+\fft12\epsilon^{\mu\nu\rho\sigma\lambda}R_{\mu\nu\alpha\beta}R_{\rho\sigma}{}^{\alpha\beta}\mathcal A_\lambda,
\end{equation}
and the second term is a mixed invariant
\begin{align}
    \alpha\, e^{-1}\mathcal L_{G^2V^2}&=-\fft3{16}\bar a_{IJ}\Bigl[G_{\mu\nu}G^{\mu\nu}V^I_{\rho\sigma}V^{J\,\rho\sigma}-6G_{\mu\nu}V^{I\,\mu\nu}G_{\rho\sigma}V^{J\,\rho\sigma}\nn\\
    &\kern5em-16GGV^IV^J+12GV^IGV^J-2\partial_\mu\varphi^I\partial^\mu\varphi^JG_{\rho\sigma}G^{\rho\sigma}\nn\\
    &\kern5em-16\partial^\mu\varphi^I\partial^\nu\varphi^JG_{\mu\rho}G_\nu{}^{\rho}+16\epsilon^{\mu\nu\rho\sigma\lambda}G_{\mu\nu}G_\rho{}^\alpha V^I_{\sigma\alpha}\partial_\lambda\varphi^J
    \Bigr].
\label{eq:mixedI}
\end{align}
Finally, the last term is a vector multiplet invariant
\begin{align}
    \fft\alpha{2\kappa^2}\, e^{-1}\mathcal L_{V^4}&=-\fft38(\bar a_{IJ}\bar a_{KL}+\bar a_{IK}\bar a_{JL}+\bar a_{IL}\bar a_{JK})\Bigl[V^I_{\mu\nu}V^{J\,\mu\nu}V^K_{\rho\sigma}V^{L\,\rho\sigma}-4V^IV^JV^KV^L\nn\\
    &\kern4em-4\partial_\mu\varphi^I\partial^\mu\varphi^J\partial_\nu\varphi^K\partial^\nu\varphi^L+4\partial_\mu\varphi^I\partial^\mu\varphi^JV^K_{\rho\sigma}V^{L\,\rho\sigma}-16\partial^\mu\varphi^I\partial^\nu\varphi^JV^K_{\mu\rho}V^L_\nu{}^\rho\Bigr]\nn\\
    &\quad-\fft34\bar a_{IJ}\bar a_{KL}\Bigl[V^I_{\mu\nu}V^{J\,\mu\nu}V^K_{\rho\sigma}V^{L\,\rho\sigma}-4V^IV^JV^KV^L-4\partial_\mu\varphi^I\partial^\mu\varphi^K\partial_\nu\varphi^J\partial^\nu\varphi^L\nn\\
    &\kern6em+4\partial_\mu\varphi^I\partial^\mu\varphi^KV^J_{\rho\sigma}V^{L\,\rho\sigma}-8(\partial^\mu\varphi^I\partial^\nu\varphi^JV^K_{\mu\rho}V^L_\nu{}^\rho+\partial^\mu\varphi^I\partial^\nu\varphi^KV^L_{\mu\rho}V^J_\nu{}^\rho)\nn\\
    &\kern6em-8\epsilon^{\mu\nu\rho\sigma\lambda}V^I_{\mu\nu}V^J_\rho{}^\alpha V^K_{\sigma\alpha}\partial_\lambda\varphi^L\Bigr].
\label{eq:vmi}
\end{align}
Note that, in order for the four-derivative vector multiplet invariants to survive in the rigid limit, we must scale the coupling $\alpha\sim\mathcal O(\kappa^2)$ before taking the $\kappa\to0$.

The above vector multiplet invariant can be written as
\begin{align}
    \tilde\alpha\,\mathcal L_{V^4}&=\Sigma_{IJKL}\Bigl[V^I_{\mu\nu}V^{J\,\mu\nu}V^K_{\rho\sigma}V^{L\,\rho\sigma}-4V^IV^JV^KV^L-4\partial_\mu\varphi^I\partial^\mu\varphi^K\partial_\nu\varphi^J\partial^\nu\varphi^L\nn\\
    &\kern4em+4\partial_\mu\varphi^I\partial^\mu\varphi^KV^J_{\rho\sigma}V^{L\,\rho\sigma}-8(\partial^\mu\varphi^I\partial^\nu\varphi^JV^K_{\mu\rho}V^L_\nu{}^\rho+\partial^\mu\varphi^I\partial^\nu\varphi^KV^L_{\mu\rho}V^J_\nu{}^\rho)\nn\\
    &\kern4em-8\epsilon^{\mu\nu\rho\sigma\lambda}V^I_{\mu\nu}V^J_\rho{}^\alpha V^K_{\sigma\alpha}\partial_\lambda\varphi^L\Bigr],
\label{eq:sigmaijkl}
\end{align}
where $\tilde\alpha=\alpha/2\kappa^2$ and where the coupling matrix $\Sigma_{IJKL}$ takes the form
\begin{equation}
    \Sigma_{IJKL}=-\fft38(\bar a_{IJ}\bar a_{KL}+\bar a_{IK}\bar a_{JL}+\bar a_{IL}\bar a_{JK})-\fft34\bar a_{IJ}\bar a_{KL}.
\end{equation}
This suggests that the vector invariant can be generalized to arbitrary couplings $\Sigma_{IJKL}$, provided it is symmetric in $(IJ)$ and $(KL)$ and symmetric under the interchange of index pairs $IJ\leftrightarrow KL$.  Note that the CP-odd term in (\ref{eq:sigmaijkl}) vanishes when $\Sigma_{IJKL}$ is symmetrized over all four indices.  We also see that, in the case of a single vector multiplet, such as in the $C_{112}$ model, there is only a single invariant that takes the form
\begin{equation}
    \mathcal L_{V^4}\sim ((V_{\mu\nu})^2)^2-4V^4-4(\partial\varphi^2)^2+4\partial\varphi^2(V_{\mu\nu})^2-16V^2_{\mu\nu}\partial^\mu\varphi\partial^\nu\varphi,
\label{eq:singlev}
\end{equation}
where we have taken canonical normalizations for the scalar and vector fields.  This agrees with the dimensional reduction of the six-dimensional super-Yang-Mills invariant~\cite{Bergshoeff:1986jm}%
\begin{equation}
    \mathcal L_{6\mathrm{D}}=(F^2)^2-4F^4.
\end{equation}

It is worth noting that the mixed invariant, (\ref{eq:mixedI}), presents a potential obstruction to the decoupling of the gravity sector at the four-derivative level.  Since $\mathcal L_{G^2V^2}\sim\mathcal O(\tilde\alpha/\kappa^2)$, and the backreaction of the vector multiplets on the gravity sector is $\mathcal O(\kappa^2)$, this can lead to $\mathcal O(1)$ corrections from the backreaction.  Nevertheless, this would not be an issue in cases where the mixed invariant is not present, such as those we saw in Section~\ref{sec:compareInv}.

\subsection{Specializing to the STU model}

In general, when there are two or more vector multiplets, the quartic vector coupling, (\ref{eq:sigmaijkl}), will involve couplings among the multiplets.  To see this explicitly, we can consider the STU model with $C_{123}=1/6$.  In the rigid limit, the three vectors, $F^I$, of the STU model can be decomposed into one graviton and two matter vectors.  To be consistent with the choice of unconstrained scalars, (\ref{eq:stuvarphi}), and for canonical normalization, we let
\begin{align}
    X^1&=e^{-\fft1{\sqrt6}\varphi_1-\fft1{\sqrt2}\varphi_2},&X^2&=e^{-\fft1{\sqrt6}\varphi_1+\fft1{\sqrt2}\varphi_2},&X^3&=e^{\fft2{\sqrt6}\varphi_1},\nn\\
    \tilde F^1&=\fft1{\sqrt3}G+\fft1{\sqrt6}V+\fft1{\sqrt2}U,&\tilde F^2&=\fft1{\sqrt3}G+\fft1{\sqrt6}V-\fft1{\sqrt2}U,&\tilde F^3&=\fft1{\sqrt3}G-\fft2{\sqrt6}V,
\end{align}
where $\tilde F^I=F^I/X^I=3X_IF^I$ (no sum).

To facilitate the expansion of the vector multiplet invariant, (\ref{eq:sigmaijkl}), we make note of the combinations
\begin{align}
    a_{IJ}\partial_\mu X^I\partial_\nu X^J&=\fft13\left(\partial_\mu\varphi_1\partial_\nu\varphi_1+\partial_\mu\varphi_2\partial_\nu\varphi_2\right),\nn\\
    a_{IJ}V_{\mu\nu}^IV_{\rho\sigma}^J&=\fft13\left(V_{\mu\nu}V_{\rho\sigma}+U_{\mu\nu}U_{\rho\sigma}\right),\nn\\
    a_{IJ}\partial_\rho X^IV_{\mu\nu}^J&=-\fft13\left(\partial_\rho\varphi_1 V_{\mu\nu}+\partial_\rho\varphi_2 U_{\mu\nu}\right).
\end{align}
For the fully symmetric combination $\Sigma_{IJKL}=\bar a_{IJ}\bar a_{KL}+\bar a_{IK}\bar a_{JL}+\bar a_{IL}\bar a_{JK}$, we find
\begin{align}
    \tilde\alpha\,\mathcal L_{\mathrm{sym}}&=\fft13\left[(V^2)^2-4V^4-4(\partial\varphi_1^2)^2+4\partial\varphi_1^2V^2-16\partial^\mu\varphi_1\partial^\nu\varphi_1V_{\mu\lambda}V_\nu{}^\lambda\right]\nn\\
    &\quad+\fft13\left[(U^2)^2-4U^4-4(\partial\varphi_2^2)^2+4\partial\varphi_2^2U^2-16\partial^\mu\varphi_1\partial^\nu\varphi_1U_{\mu\lambda}U_\nu{}^\lambda\right]\nn\\
    &\quad+\fft23\Bigl[V^2U^2+2(V\cdot U)^2-8VVUU-4VUVU-4\partial\varphi_1^2\partial\varphi_2^2-8(\partial\varphi_1\cdot\partial\varphi_2)^2\nn\\
    &\kern3em+2\partial\varphi_1^2U^2+2\partial\varphi_2^2V^2+8\partial\varphi_1\cdot\partial\varphi_2V\cdot U-8\partial^\mu\varphi_1\partial^\nu\varphi_1U^2_{\mu\nu}-8\partial^\mu\varphi_2\partial^\nu\varphi_2V^2_{\mu\nu}\nn\\
    &\kern3em-16\partial^\mu\varphi_1\partial^\nu\varphi_2(U_{\mu\lambda}V_\nu{}^\lambda+V_{\mu\lambda}U_\nu{}^\lambda)\Bigr],
\label{eq:sv4inv}
\end{align}
and for the pairwise combination $\Sigma_{IJKL}=\bar a_{IJ}\bar a_{KL}$, we find
\begin{align}
    \tilde\alpha\,\mathcal L_{\mathrm{pair}}&=\fft19\left[(V^2)^2-4V^4-4(\partial\varphi_1^2)^2+4\partial\varphi_1^2V^2-16\partial^\mu\varphi_1\partial^\nu\varphi_1V_{\mu\lambda}V_\nu{}^\lambda\right]\nn\\
    &\quad+\fft19\left[(U^2)^2-4U^4-4(\partial\varphi_2^2)^2+4\partial\varphi_2^2U^2-16\partial^\mu\varphi_1\partial^\nu\varphi_1U_{\mu\lambda}U_\nu{}^\lambda\right]\nn\\
    &\quad+\fft29\Bigl[V^2U^2-4VVUU-4(\partial\varphi_1\cdot\partial\varphi_2)^2+4\partial\varphi_1\cdot\partial\varphi_2V\cdot U\nn\\
    &\kern3em-4\partial^\mu\varphi_1\partial^\nu\varphi_1U^2_{\mu\nu}-4\partial^\mu\varphi_2\partial^\nu\varphi_2V^2_{\mu\nu}-8\partial^\mu\varphi_1\partial^\nu\varphi_2U_{\mu\lambda}V_\nu{}^\lambda\nn\\
    &\kern3em+4\epsilon^{\mu\nu\rho\sigma\lambda}(V_{\mu\nu}V_{\rho\alpha}U_\sigma{}^\alpha\partial_\lambda\varphi_2+U_{\mu\nu}U_{\rho\alpha}V_\sigma{}^\alpha\partial_\lambda\varphi_1)\Bigr].
\label{eq:pv4inv}
\end{align}
We conjecture that the symmetric and pairwise combinations are independent superinvariants by themselves.  We also see that the first two lines of either (\ref{eq:sv4inv}) or (\ref{eq:pv4inv}) are just the single vector invariant, (\ref{eq:singlev}).

\section{Discussion}\label{sec:disc2}
We have made significant progress in classifying four-derivative superinvariants in five-dimensional $\mathcal N=2$ supergravity coupled to vector multiplets. In particular, we have found a new vector invariant for the $C_{112}$ model and two new vector invariants for the STU model. Notably, all of these vector invariants consist exclusively of CP-even terms. Moreover, these superinvariants do not affect the static three-charge BPS solution, and they have a rigid limit to $\mathcal N=2$ super-Yang-Mills invariants. Nevertheless, our treatment is far from exhaustive, and there is still much work to do in classifying five-dimensional $\mathcal N=2$ superinvariants.

It would be interesting to see if such vector invariants could be constructed in general using a bottom-up construction, such as superconformal tensor calculus. An $F^4$-type invariant was constructed using superconformal tensor calculus in~\cite{Ozkan:2016csy}, but this seems to be zero after appropriate field redefinitions in the ungauged case. The supersymmetrization of the Ricci tensor squared action in the standard Weyl multiplet has only been written down for the case of pure minimal five-dimensional supergravity~\cite{Gold:2023ykx}; however, in principle, this admits an extension to the case of an arbitrary number of vector multiplets. It would be interesting to see if this reduces to one of the vector multiplets found in the $C_{112}$ and STU cases. More generally, such a bottom-up construction would be useful to extend the vector superinvariants we have found to an arbitrary number of vector multiplets.

One could also treat the heterotic reduction more generally. In particular, reduction from ten to five dimensions leads to $\mathcal N=4$ supergravity coupled to five vector multiplets, which leads to ten vector fields $F^{(\pm)\,a}$. From an $\mathcal N=2$ perspective, four of the $F^{(-)\,a}$ correspond to the two gravitino multiplets, while the fifth $F^{(-)\,a}$ combines with the (dualized) NS flux to give the graviphoton in the gravity multiplet and a vector multiplet. The remaining $F^{(+)\,a}$, along with five of the $P^{(+-)\,ab}$, form five $\mathcal N=2$ vector multiplets. The remaining scalars live in hypermultiplets. Thus, truncating the gravitini and hypermultiplets would leave us with $\mathcal N=2$ supergravity coupled to six vector multiplets, which generalizes the STU case studied in this paper. As shown in~\cite{Cai:2025yyv}, this is consistent with $\mathcal N=2$ supersymmetry, at least at the two-derivative level. This would then also truncate to the case of $1\le n_v\le 6$ $\mathcal N=2$ vector multiplets. At the very least, this could be compared with the $C^2+\frac{1}{6}R^2$ invariant constructed using the standard Weyl multiplet,~\eqref{eq: Cassani Lagrangian}. Of course, this would correspond to a very particular choice of cubic prepotential, but it would be interesting to see if this leads to new vector invariants. Even more generally, one could also keep the heterotic gauge fields in the reduction, leading to $\mathcal N=2$ supergravity coupled to 22 vector multiplets.

Finally, it would be useful to extend our results to gauged supergravity, as this would apply to precision holography. Turning on the gauging in the heterotic theory would require reduction on a curved internal manifold, for which the only known four-derivative example is the reduction on $S^3$~\cite{Liu:2023fmv},%
\footnote{Note that Ref.~\cite{Liu:2023fmv} truncates the half-maximal vector multiplets that arise in the reduction.}
but it is not clear if this particular example admits a consistent truncation to the STU model. Fortunately, for the standard Weyl multiplet construction, the gauged Weyl-squared invariant is known~\cite{Hanaki:2006pj}, and, for the dilaton Weyl multiplet construction, the gauged Weyl-squared and Ricci-squared actions are known~\cite{Gold:2023ymc}. In particular, the dilaton Weyl Ricci-squared invariant should lead to an $F^4$ invariant in gauged $C_{112}$ supergravity, corresponding to the gauged version of Eq.~\eqref{eq:DWRic2ein}. It would be interesting to see if the resulting vector invariant affects BPS black hole solutions in AdS$_5$, as this could impact their thermodynamics and holographic matching with the superconformal index \cite{Bobev:2022bjm,Cassani:2022lrk}.

\section*{Acknowledgements}
SJ was supported in part by a Leinweber Graduate Fellowship. YP and RJS are supported by National Key Research and Development Program No.~2022YFE0134300 and the National Natural Science Foundation of China (NSFC) under Grants No.~12575076 and No.~12247103.

\appendix

\section{Field redefinitions for the standard Weyl multiplet construction}\label{app:fieldReds}

Since we are comparing multiple four-derivative gravitational invariants, it is important that we work in a uniform field redefinition framework.  Our choice is to use a Riemann basis for the gravitational couplings and to avoid any higher than first derivatives of the fields.  In particular, we will use field redefinitions to replace the Gauss-Bonnet combination $\chi_{\mathrm{GB}}$ and the Weyl tensor with corresponding expressions written using the Riemann and Ricci tensors.  The Ricci terms can then be eliminated through the use of the Einstein equation.  Likewise, terms such as $\nabla F$ and $\nabla\nabla X$ will be eliminated through a combination of integration by parts and the equations of motion.

In this Appendix, we focus on the Weyl-squared invariant, (\ref{eq: Cassani Lagrangian}), constructed using the standard Weyl multiplet~\cite{Cassani:2024tvk}, which we repeat here for convenience
\begin{align}
\mathcal{L}^{\mathrm{SW}}_{C^2}= & \,\lambda_M X^M \mathcal{X}_{\mathrm{GB}}+D_{I J} C_{\mu \nu \rho \sigma} F^{I \mu \nu} F^{J \rho \sigma}+E_{I J K L} F_{\mu \nu}^I F^{J \mu \nu} F_{\rho \sigma}^K F^{L \rho \sigma}\nn \\
& +\widetilde{E}_{I J K L} F_{\mu \nu}^I F^{J \nu \rho} F_{\rho \sigma}^K F^{L \sigma \mu}+I_{I J K L} \partial_\mu X^I \partial^\mu X^J \partial_\nu X^K \partial^\nu X^L+H_{I J K L} \partial_\mu X^I \partial^\mu X^J F_{\rho \sigma}^K F^{L \rho \sigma} \nn\\
& +\widetilde{H}_{I J K L} \partial_\mu X^I \partial^\nu X^J F^{K \mu \rho} F_{\nu \rho}^L-6 X_I X_J \lambda_K F^{I \mu \alpha} F^{J \nu}{ }_\alpha \nabla_\nu \nabla_\mu X^K\nn \\
& +\frac{3}{4} \lambda_{[I} X_{J]} X_K \epsilon^{\mu \nu \rho \sigma \lambda} \nabla_\alpha F_{\mu \nu}^I F_{\rho \sigma}^J F_\lambda^{K \alpha}+W_{I J K L} \epsilon^{\mu \nu \rho \sigma \lambda} F_{\mu \nu}^I F_\rho^{J \alpha} F_{\sigma \alpha}^K \partial_\lambda X^L\nn \\
& +\frac{1}{2} \lambda_I \epsilon^{\mu \nu \rho \sigma \lambda} R_{\mu \nu \alpha \beta} R_{\rho \sigma}{ }^{\alpha \beta} A_\lambda^I.
\label{eq: Cassani Lagrangian v2}
\end{align}
The various tensors, $D_{IJ}$, $E_{IJKL}$,\ldots, are constructed from the constrained scalars $X^I$ and take the form~\cite{Cassani:2024tvk}
\begin{align}
D_{I J}= &\, 3 \lambda_I X_J-\frac{9}{2} \lambda_M X^M X_I X_J, \nn\\
E_{I J K L}= & \,\lambda_M X^M\left(-\frac{27}{16} X_I X_J X_K X_L-\frac{9}{8} a_{I J} a_{K L}+\frac{39}{16} a_{I J} X_K X_L-\frac{3}{4} a_{I K} a_{J L}+\frac{9}{4} a_{I K} X_J X_L\right) \nn\\
& -\frac{3}{4} a_{J[I} \lambda_{L]} X_K-\frac{9}{8} \lambda_I X_J X_K X_L,\nn \\
\widetilde{E}_{I J K L}= &\, \lambda_M X^M\left(\frac{81}{8} X_I X_J X_K X_L+6 a_{I J} a_{K L}+\frac{3}{2} a_{I K} a_{J L}-9 X_I X_J a_{K L}-\frac{9}{2} X_I X_K a_{J L}\right)\nn \\
& -\frac{9}{4} X_I X_J X_K \lambda_L-\frac{3}{4} a_{I K} X_J \lambda_L, \nn\\
I_{I J K L}= & \,\lambda_M X^M\left(\frac{3}{2} a_{I J} a_{K L}+6 a_{K(I} a_{J) L}\right), \nn\\
H_{I J K L}= &\, -\frac{3}{2} \lambda_M X^M a_{I J} a_{K L}-6 \lambda_M X^M a_{I K} a_{J L}+\frac{45}{8} \lambda_M X^M a_{I J} X_K X_L-\frac{9}{4} a_{I J} X_K \lambda_L, \nn\\
\widetilde{H}_{I J K L}= &\, \lambda_M X^M\left(12 a_{I L} a_{J K}+6 a_{I K} a_{J L}+12 a_{I J} a_{K L}-9 a_{I J} X_K X_L\right), \nn\\
W_{I J K L}= & \,-\frac{21}{4} \lambda_J a_{I L} X_K-\frac{3}{4} \lambda_J X_I a_{K L}+3 \lambda_M X^M\left(2 a_{I J} a_{K L}-3 X_I X_J a_{K L}\right) .
\label{eq:tensors}
\end{align}
We see that there are four terms in (\ref{eq: Cassani Lagrangian v2}) that need to be addressed by field redefinitions.  These are the first two terms that involve $\chi_{\mathrm{GB}}$ and $C_{\mu\nu\rho\sigma}$, respectively, and the second derivative terms $\nabla\nabla X$ and $\nabla F$.

Using the two-derivative Einstein equation, the first term, $\lambda_MX^M\mathcal{X}_\mathrm{GB}$ can be replaced by $\lambda_MX^MR_{\mu\nu\rho\sigma}^2$ along with shifting
\begin{align}
    E_{IJKL}&\to E_{IJKL}+\fft{29}{16}\lambda_MX^Ma_{IJ}a_{KL},\nn\\
    \widetilde E_{IJKL}&\to\widetilde E_{IJKL}-9\lambda_MX^Ma_{IJ}a_{KL},\nn\\
    I_{IJKL}&\to I_{IJKL}+\lambda_MX^M\left(-9a_{IK}a_{JL}+\fft94a_{IJ}a_{KL}\right),\nn\\
    H_{IJKL}&\to H_{IJKL}+\fft{15}4\lambda_MX^Ma_{IJ}a_{KL},\nn\\
    \widetilde H_{IJKL}&\to \widetilde H_{IJKL}-18\lambda_MX^Ma_{IJ}a_{KL}.
\label{eq:CFR1}
\end{align}
Similarly, the second term, $D_{IJ}C_{\mu\nu\rho\sigma}F^{I\,\mu\nu}F^{J\,\rho\sigma}$, reduces to $D_{IJ}R_{\mu\nu\rho\sigma}F^{I\,\mu\nu}F^{J\,\rho\sigma}$, along with shifting
\begin{align}
    E_{IJKL}&\to E_{IJKL}+\fft38D_{IJ}a_{KL},\nn\\
    \widetilde E_{IJKL}&\to \widetilde E_{IJKL}-2D_{IJ}a_{KL},\nn\\
    H_{IJKL}&\to H_{IJKL}+\fft14a_{IJ}D_{KL},\nn\\
    \widetilde H_{IJKL}&\to\widetilde H_{IJKL}-2a_{IJ}D_{KL}.
\label{eq:CFR2}
\end{align}

For the $\nabla F$ and $\nabla\nabla X$ terms, we first use integration by parts and Bianchi identities to rewrite
\begin{align}
    \lambda_{[I} X_{J]} X_K\epsilon^{\mu\nu\rho\sigma\lambda}\nabla_\alpha F_{\mu\nu}^IF_{\rho\sigma}^JF_{\lambda\alpha}^K&\to\epsilon^{\mu\nu\rho\sigma\lambda}\Bigg[\frac{1}{2}\lambda_I X_J X_K \nabla^\alpha F_{\nu\alpha}^I F_{\rho\sigma}^J F_{\mu\lambda}^K+\frac{1}{2}\lambda_I\partial_\alpha\qty(X_JX_K)F^I_{\nu\alpha}F^J_{\rho\sigma}F^K_{\mu\lambda}\nn\\
    &\qquad\qquad+2\lambda_I\partial_\mu\qty(X_JX_K)F^I_{\nu\alpha}F^J_{\rho\sigma}F^K_{\lambda\alpha}+\frac{1}{2}\lambda_J\partial_\alpha\qty(X_IX_K)F^I_{\mu\nu}F^J_{\rho\sigma}F^K_{\lambda\alpha}\nn\\
    &\qquad\qquad+\frac{1}{2}\lambda_JX_IX_KF^I_{\mu\nu}F^J_{\rho\sigma}\nabla^\alpha F^K_{\lambda\alpha}\Bigg],
\end{align}
and
\begin{align}
    X_IX_J\lambda_KF^I_{\mu\alpha}F^J_{\nu\alpha}\nabla_\nu\partial_\mu X^K&\to \frac{1}{4}X_IX_J\lambda_K (F^2)^{IJ}\Box X^K-\partial_\nu\qty(X_IX_J)\lambda_K F^I_{\mu\alpha}F^J_{\nu\alpha}\partial^\mu X^K\nn\\
    &\quad-X_IX_J\lambda_KF^I_{\mu\alpha}\nabla^\nu F_{\nu\alpha}^J\partial^\mu X^K+\frac{1}{4}\partial_\mu\qty(X_IX_J)\lambda_K (F^2)^{IJ}\partial^\mu X^K.
\end{align}
These expressions can be further manipulated using the scalar and gauge equations of motion.

The scalar equations of motion for the $X^I$ must be treated with some care due to the prepotential constraint $\mathcal V=1$. These were obtained in~\cite{Cremonini:2008tw} and read
\begin{align}
    0&=\qty(\delta^J_I-X_IX^J)\qty[C_{JKL}\qty(2X^K\Box X^L+\partial_\mu X^K\partial^\mu X^L)-\frac{1}{2}\qty(\mathcal C_{JK}X_L-C_{JKL})F^K_{\mu\nu}F^L_{\mu\nu}],
\end{align}
where $\mathcal C_{IJ}=6C_{IJK}X^K$.  Contracting the free index with $\mathcal C^{IJ}$ and making use of the identity
\begin{equation}
    0=\frac{1}{3}\Box\mathcal V=C_{IJK}X^I\qty(X^J\Box X^K+2\partial_\mu X^J\partial^\mu X^K),
\end{equation}
we find that
\begin{equation}
    \Box X^I=(a^{IJ}-3X^IX^J)C_{JKL}\partial_\mu X^K\partial^\mu X^L-\fft12(a^{IJ}-X^IX^J)(3a_{JK}X_L+C_{JKL})F_{\mu\nu}^KF^{L\,\mu\nu},
\end{equation}
where the inverse of $a_{IJ}$ is given by
\begin{equation}
    a^{IJ}=\frac{3}{2}X^IX^J-3\mathcal C^{IJ},
\end{equation}
and $\mathcal C^{IJ}$ is the inverse of $\mathcal C_{IJ}$.
The gauge field equations of motion are
\begin{equation}
    \nabla^\nu\qty(a_{IJ}F^J_{\nu\mu})=-\frac{1}{4}\mathcal C_{IJK}\epsilon_\mu{}^{\nu\rho\sigma\lambda}F^J_{\nu\rho}F^K_{\rho\sigma},
\end{equation}
and may be brought into the form
\begin{equation}
    \nabla^\lambda F_{\mu\lambda}^I=2a^{IJ}(3a_{L(J}X_{K)}+C_{JKL})F_{\mu\lambda}^K\partial^\lambda X^L+\fft14a^{IJ}C_{JKL}\epsilon_\mu{}^{\nu\rho\sigma\lambda}F_{\nu\rho}^KF_{\sigma\lambda}^L.
\end{equation}

With these manipulations in mind, we see that the term $-6X_IX_J\lambda_KF^{I\,\mu\alpha}F^{J\,\nu}{}_\alpha\nabla_\nu\nabla_\mu X^K$ can be replaced by the shifts
\begin{align}
    E_{IJKL}&\to E_{IJKL}+\fft34X_IX_J\lambda_M(a^{MN}-X^MX^N)(3a_{NK}X_L+C_{NKL}),\nn\\
    H_{IJKL}&\to H_{IJKL}+3\lambda_Ia_{JL}X_K-\fft32X_KX_L\lambda_M(a^{MN}-3X^MX^N)C_{NIJ},\nn\\
    \widetilde H_{IJKL}&\to\widetilde H_{IJKL}-6\lambda_Ja_{IL}X_K-6\lambda_Ja_{IK}X_L+12\lambda_Ia_{JL}X_K,
\label{eq:CFR3}
\end{align}
along with generating a new term of the form $\widetilde W_{IJKL} \epsilon^{\mu\nu\rho\sigma\lambda} F_{\mu\nu}^I F_{\rho\sigma}^J F_{\lambda\alpha}^K \partial^\alpha X^L$, where
\begin{equation}
    \widetilde W_{IJKL}=\fft32\lambda_LC_{IJM}X^MX_K.
\label{eq:wtW}
\end{equation}
In addition, the term $\fft34\lambda_{[I}X_{J]}X_K\epsilon^{\mu\nu\rho\sigma\lambda}\nabla_\alpha F_{\mu\nu}^IF_{\rho\sigma}^JF^K_\lambda{}^\alpha$ can be replaced by the shifts
\begin{align}
    E_{IJKL}&\to E_{IJKL}+\fft34\lambda_Ma^{MN}C_{NJL}X_IX_K+\fft38\lambda_K(a_{JL}-3X_JX_L)X_I,\nn\\
    \widetilde E_{IJKL}&\to \widetilde E_{IJKL}-\fft32\lambda_Ma^{MN}C_{NJL}X_IX_K-\fft34\lambda_K(a_{JL}-3X_JX_L)X_I,\nn\\
    W_{IJKL}&\to W_{IJKL}-\fft32\lambda_JX_Ka_{IL}-\fft32\lambda_JX_Ia_{KL},\nn\\
    \widetilde W_{IJKL}&\to\widetilde W_{IJKL}+\fft34\lambda_KX_Ia_{JL}+\fft38\lambda_JX_Ia_{KL}-\fft98\lambda_LX_IX_JX_K-\fft98\lambda_MX^MX_IX_Ja_{KL}\nn\\
    &\quad-\fft34\lambda_Ma^{MN}C_{NKL}X_IX_J-\fft38\lambda_IX_Ka_{JL}.
\label{eq:CFR4}
\end{align}

Combining the field redefinitions (\ref{eq:CFR1}), (\ref{eq:CFR2}), (\ref{eq:CFR3}), (\ref{eq:wtW}) and (\ref{eq:CFR4}), we may bring the Lagrangian \eqref{eq: Cassani Lagrangian v2} into the form
\begin{align}
    \mathcal{L}^{\mathrm{SW}}_{C^2}= & \,\lambda_M X^M R_{\mu\nu\rho\sigma}^2 +D_{I J} R_{\mu \nu \rho \sigma} F^{I \mu \nu} F^{J \rho \sigma}+I'_{I J K L} \partial_\mu X^I \partial^\mu X^J \partial_\nu X^K \partial^\nu X^L\nn \\
    &+ E'_{I J K L} F_{\mu \nu}^I F^{J \mu \nu} F_{\rho \sigma}^K F^{L \rho \sigma}  +\widetilde{E}'_{I J K L} F_{\mu \nu}^I F^{J \nu \rho} F_{\rho \sigma}^K F^{L \sigma \mu}+ \nn\\
    & + H'_{I J K L} \partial_\mu X^I \partial^\mu X^J F_{\rho \sigma}^K F^{L \rho \sigma} + \widetilde{H}'_{I J K L} \partial_\mu X^I \partial^\nu X^J F^{K \mu \rho} F_{\nu \rho}^L\nn \\
    & +W'_{I J K L} \epsilon^{\mu \nu \rho \sigma \lambda} F_{\mu \nu}^I F_\rho^{J \alpha} F_{\sigma \alpha}^K \partial_\lambda X^L+ \widetilde W_{IJKL} \epsilon^{\mu\nu\rho\sigma\lambda} F_{\mu\nu}^I F_{\rho\sigma}^J F_{\lambda\alpha}^K \partial^\alpha X^L\nn\\
    &+\frac{1}{2} \lambda_I \epsilon^{\mu \nu \rho \sigma \lambda} R_{\mu \nu \alpha \beta} R_{\rho \sigma}{ }^{\alpha \beta} A_\lambda^I,\label{eq:CassaniFinalForm 1}
\end{align}
where
\begin{align}
    E'_{IJKL}&=E_{IJKL} + \dfrac{1}{8}\lambda_M X^M \left(\dfrac{29}{2}a_{IJ}a_{KL} - 27 X_IX_JX_KX_L + 3 X_IX_J a_{KL} \right) +\dfrac{3}{8}D_{IJ}a_{KL}\nn\\
    &\quad  + \dfrac{3}{8} a_{IK}\lambda_{J} X_{L}+ \dfrac{9}{8}X_IX_J\lambda_KX_L + \dfrac{3}{4} \lambda_N a^{MN}\qty(X_IX_J C_{KLM} + X_JX_LC_{IKM}),\nn\\
    \widetilde E'_{IJKL} &=\widetilde E_{IJKL} - 9 \lambda_MX^M a_{IJ}a_{KL} - 2 D_{IJ}a_{KL} - \dfrac{3}{2} \lambda_M a^{MN} C_{IKN} X_J X_L \nn\\
    &\quad + \dfrac{9}{4} X_I \lambda_J X_K X_L - \dfrac{3}{4} a_{IK}\lambda_J X_L, \nn\\
    I_{IJKL}'&= I_{IJKL} - 9 \lambda_MX^M \left( a_{IK}a_{JL} - \dfrac{1}{4} a_{IJ} a_{KL} \right),\nn\\
    H_{IJKL}'&=H_{IJKL} + \dfrac{3}{4}\lambda_MX^M \left(5a_{IJ}a_{KL} -3 a_{IJ}X_KX_L \right) + \dfrac{1}{4} a_{IJ}D_{KL}\nn\\
    &\quad+3\lambda_Ia_{JL}X_K  - \dfrac{3}{2} \lambda_M a^{MN} C_{NIJ}X_KX_L,\nn\\
    \widetilde H_{IJKL}'&=\widetilde H_{IJKL} - 18 \lambda_M X^M a_{IJ}a_{KL} - 2 a_{IJ} D_{KL} - 6 a_{IL}\lambda_J X_K - 6 a_{IK}\lambda_J X_L
    + 12 \lambda_I a_{JL} X_K,\nn\\
    W_{IJKL}'&= W_{IJKL} -3 X_{(I}a_{K)L}\lambda_J,\nn\\
    \widetilde W_{IJKL}'&=-\dfrac{9}{8} \lambda_M X^M X_I X_J a_{KL} - \dfrac{3}{4} a_{IJ} X_K \lambda_L + \dfrac{3}{4} a_{IL} X_J \lambda_K + \dfrac{3}{4} X_{[I} a_{K]L} \lambda_J\nn \\
    &\quad+ \dfrac{9}{8} X_I X_J X_K \lambda_L - \dfrac{3}{4} \lambda_M a^{MN}  X_I X_J  C_{NKL}.
\end{align}

Finally, while there are two forms of $\epsilon FFF\partial X$, parametrized by $W_{IJKL}$ and $\tilde W_{IJKL}$, the latter terms can be rewritten in terms of the former by using the five-dimensional identity
\begin{equation}
    \epsilon^{\mu\nu\rho\sigma\lambda}F_{\mu\nu}G_{\rho\sigma}H_{\lambda\alpha}\partial^\alpha X=2\epsilon_{\mu\nu\rho\sigma\lambda}(F_{\mu\nu}H_{\rho\alpha}G_\sigma{}^\alpha+G_{\mu\nu}H_{\rho\alpha}F_\sigma{}^\alpha)\partial_\lambda X.
\end{equation}
This allows us to replace
\begin{equation}
    \widetilde W_{IJKL}\quad\to\quad W_{IJKL}=2(\widetilde W_{IKJL}+\widetilde W_{KIJL}).
\label{eq:epsid}
\end{equation}
This gives rise to the final field redefined form of the Lagrangian, (\ref{eq:CassaniFinalForm}), along with the final expressions for the tensors, (\ref{eq:cfftensors}).

\section{Field redefinitions for the dilaton Weyl multiplet construction}
\label{app:dwfield}

At the two-derivative level, the Lagrangian for the dilaton Weyl construction is given in the string frame by (\ref{eq:DWstringFrame}) and the Einstein frame by (\ref{eq:DWeinTwoDeriv}).  The corresponding string frame equations of motion are
\begin{align}
    \mathcal E_{g,\alpha\beta}&=R_{\alpha\beta}-G^2_{\alpha\beta}-\frac{1}{2}\qty(H^2_{\alpha\beta}-\frac{1}{2}H^2g_{\alpha\beta})+2\nabla_\alpha\varphi\nabla_\beta\varphi,\nn\\
    \mathcal E_{G,\alpha}&=e^{2\varphi}\nabla^\beta\qty(e^{-2\varphi}G_{\alpha\beta})+\frac{1}{4}e^{2\varphi}\epsilon_\alpha{}^{\beta\gamma\delta\epsilon}G_{\beta\gamma}H_{\delta\epsilon},\nn\\
    \mathcal E_{H,\alpha}&=e^{-2\varphi}\nabla^\beta\qty(e^{2\varphi}H_{\alpha\beta})+\frac{1}{4}e^{-2\varphi}\epsilon_\alpha{}^{\beta\gamma\delta\epsilon}G_{\beta\gamma}G_{\delta\epsilon},\nn\\
    \mathcal E_{\varphi}&=\frac{1}{4}R+\Box\varphi-(\partial\varphi)^2-\frac{1}{8}G^2+\frac{1}{16}H^2,
\label{eq:DWeoms}
\end{align}
while the Einstein frame equations of motion are
\begin{align}
    \mathcal E_{g,\alpha\beta}^\mathrm{E}&=R_{\alpha\beta}-\frac{4}{3}\partial_\alpha\varphi\partial_\beta\varphi-e^{-4\varphi/3}G^2_{\alpha\beta}-\frac{1}{2}e^{8\varphi/3}H^2_{\alpha\beta}-\frac{1}{2}g_{\mu\nu}e^{-1}\mathcal L_{\partial^2}^E,\nn\\
    \mathcal E_{G,\alpha}^\mathrm{E}&=e^{4\varphi/3}\nabla^\beta\qty(e^{-4\varphi/3}G_{\alpha\beta})+\frac{1}{4}e^{4\varphi/3}\epsilon_\alpha{}^{\beta\gamma\delta\epsilon}G_{\beta\gamma}H_{\delta\epsilon},\nn\\
    \mathcal E_{H,\alpha}^\mathrm{E}&=e^{-8\varphi/3}\nabla^\beta\qty(e^{8\varphi/3}H_{\alpha\beta})+\frac{1}{4}e^{-8\varphi/3}\epsilon_\alpha{}^{\beta\gamma\delta\epsilon}G_{\beta\gamma}G_{\delta\epsilon},\nn\\
    \mathcal E_{\varphi}^\mathrm{E}&=\Box\varphi+\frac{1}{4}e^{-4\varphi/3}G^2-\frac{1}{4}e^{8\varphi/3}H^2.
\end{align}
These two-derivative equations of motion are needed when performing the field redefinitions.

The four-derivative invariant, (\ref{eq:DWc2+16R2orig}), contains several terms involving derivatives of the field strengths $G$ and $H$.  These can be removed using Bianchi identities and integration by parts, which allows for the replacement
\begin{align}
    G^2_{\alpha\beta}\nabla^{\alpha}\nabla^\beta\varphi&\to-\nabla^\alpha G_{\alpha\gamma}G_\beta{}^\gamma\partial^\beta\varphi+\frac{1}{4}G^2\Box\varphi,\nn\\
    e^{4\varphi}H^2_{\alpha\beta}\nabla^{\alpha}\nabla^\beta\varphi&\to-\nabla^\alpha \qty(e^{4\varphi}H_{\alpha\gamma})H_\beta{}^\gamma\partial^\beta\varphi+e^{4\varphi}H^2\qty[(\partial\varphi)^2+\frac{1}{4}\Box\varphi],\nn\\
    e^{2\varphi}H_{\alpha\gamma}G_{\beta}{}^\gamma\nabla^{\alpha}\nabla^\beta\varphi&\to-\frac{1}{2}\nabla_\alpha\qty(e^{2\varphi}H^{\alpha\gamma})G_{\beta\gamma}\partial^\beta\varphi-\frac{1}{2}\nabla_\alpha\qty(G^{\alpha\gamma})e^{2\varphi}H_{\beta\gamma}\partial^\beta\varphi\nn\\
    &\quad\,+e^{2\varphi}\qty[-G_{\alpha\gamma}H_{\beta}{}^{\gamma}\partial^\alpha\varphi\partial^\beta\varphi+\frac{1}{2}(G\cdot H)(\partial\varphi)^2+\frac{1}{4}(G\cdot H)\Box\varphi],\nn\\
    (\nabla\nabla\varphi)^2&\to -R_{\alpha\beta}\partial^\alpha\varphi\partial^\beta\varphi+\qty(\Box\varphi)^2,\nn\\
    \nabla_{\alpha}\nabla_\beta\varphi\partial^\alpha\varphi\partial^\beta\varphi&\to-\frac{1}{2}(\partial\varphi)^2\Box\varphi,\nn\\
    \qty(\nabla G)^2&\to R_{\alpha\beta\gamma\delta}G^{\alpha\beta}G^{\gamma\delta}-2R^{\alpha\beta}G^2_{\alpha\beta}+2(\nabla^\alpha G_{\alpha\beta})^2,\nn\\
    \qty(\nabla(e^{2\varphi}H))^2&\to e^{4\varphi}\qty[R_{\alpha\beta\gamma\delta}H^{\alpha\beta}H^{\gamma\delta}-2R^{\alpha\beta}H^2_{\alpha\beta}+4H^2_{\alpha\beta}\nabla^\alpha\nabla^\beta\varphi-H^2\Box\varphi]\nn\\
    &\quad\,+2\qty(\nabla^\alpha (e^{2\varphi}H_{\alpha\beta}))^2-4e^{2\varphi}H_{\beta}{}^{\gamma}\nabla^\alpha H_{\gamma\alpha}\partial^\beta\varphi,\label{eq:DWsimpCPeven}
\end{align}
and
\begin{align}
    e^{2\varphi}\epsilon^{\alpha\beta\gamma\delta\epsilon}H_{\beta\omega}\nabla_\alpha \qty(e^{2\varphi}H_\gamma{}^\omega) G_{\delta\epsilon}&\to e^{4\varphi}\epsilon^{\alpha\beta\gamma\delta\epsilon}H_{\alpha\beta}\bigg[H_{\gamma\delta}G_{\epsilon\omega}\partial^\omega\varphi-G_{\gamma\delta}H_{\epsilon\omega}\partial^\omega\varphi+4H_{\gamma\omega}G_\delta{}^\omega\partial_\epsilon\varphi\nn\\
    &\kern8em\left.-\frac{1}{2}G_{\gamma\delta}e^{-2\varphi}\nabla^\omega\qty(e^{2\varphi}H_{\epsilon\omega})+\frac{1}{4}H_{\gamma\delta}\nabla^\omega G_{\epsilon\omega}\right].\label{eq:DWsimpCPodd}
\end{align}
Using \eqref{eq:DWsimpCPeven} and \eqref{eq:DWsimpCPodd} along with the string frame equations of motion, \eqref{eq:DWeoms}, we can field redefine \eqref{eq:DWc2+16R2orig} to the simplified form
\begin{align}
    e^{-1}\mathcal L_{C^2+\frac{1}{6}R^2}^\mathrm{DW}&=-\frac{1}{4}R_{\alpha\beta\gamma\delta}^2-\frac{1}{4}e^{2\varphi}R_{\alpha\beta\gamma\delta}G^{\alpha\beta}H^{\gamma\delta}+\frac{1}{8}e^{4\varphi}R_{\alpha\beta\gamma\delta}H^{\alpha\beta}H^{\gamma\delta}-\frac{1}{8}\epsilon^{\alpha\beta\gamma\delta\epsilon}R_{\alpha\beta\lambda\tau}R_{\gamma\delta}{}^{\lambda\tau}C_\epsilon\nn\\
    &\quad+\frac{5}{6}G^4-\frac{1}{48}\qty(G^2)^2+\frac{1}{12}e^{2\varphi}GGGH+\frac{1}{24}e^{2\varphi}G^2G\cdot H-\frac{1}{12}e^{4\varphi}GHGH\nn\\
    &\quad+\frac{1}{3}e^{4\varphi}GGHH+\frac{1}{24}e^{4\varphi}(G\cdot H)^2-\frac{17}{48}e^{4\varphi}G^2H^2-\frac{1}{6}e^{6\varphi}GHHH\nn\\
    &\quad+\frac{1}{24}e^{6\varphi}(G\cdot H)H^2+\frac{5}{96}e^{8\varphi}H^4+\frac{5}{48}e^{8\varphi}\qty(H^2)^2-\frac{4}{3}G^2(\partial\varphi)^2+\frac{4}{3}G^2_{\alpha\beta}\partial^\alpha\varphi\partial^\beta\varphi\nn\\
    &\quad+\frac{1}{3}e^{2\varphi}(G\cdot H)(\partial\varphi)^2-\frac{4}{3}e^{2\varphi}G_{\alpha\gamma}H_{\beta}{}^\gamma\partial^\alpha\varphi\partial^\beta\varphi+\frac{7}{6}e^{4\varphi}H^2(\partial\varphi)^2-\frac{2}{3}e^{4\varphi}H^2_{\alpha\beta}\partial^\alpha\varphi\partial^\beta\varphi\nn\\
    &\quad+\frac{4}{3}(\partial\varphi)^4\nn\\
    &\quad+\epsilon^{\alpha\beta\gamma\delta\epsilon}\left[-\frac{1}{6}e^{2\varphi}G_{\alpha\beta}G_{\gamma\delta}H_{\epsilon\omega}\partial^\omega\varphi-\frac{1}{3}e^{2\varphi}G_{\alpha\beta}H_{\gamma\delta}G_{\epsilon\omega}\partial^\omega\varphi+\frac{5}{24}e^{4\varphi}G_{\alpha\beta}H_{\gamma\delta}H_{\epsilon\omega}\partial^\omega\varphi\right.\nn\\
    &\kern4em\quad\left.-\frac{1}{48}e^{4\varphi}H_{\alpha\beta}H_{\gamma\delta}G_{\epsilon\omega}\partial^\omega\varphi+\frac{1}{2}e^{4\varphi}\partial_\alpha\varphi G_{\beta\omega}H_{\gamma}{}^\omega H_{\delta\epsilon}\right].
\end{align}

Finally, we Weyl scale this Lagrangian to the Einstein frame using (\ref{eq:DWweylRescale}).  Note that the Weyl scaling of the Riemann tensor introduces terms involving $\partial_\alpha\varphi\partial_\beta\varphi$ and $\nabla_\alpha\nabla_\beta\varphi$.  The latter can be removed using the integration by parts identities
\begin{align}
    e^{2\varphi/3}\nabla_\alpha\nabla_\beta\varphi\partial^\alpha\varphi\partial^\beta\varphi&\to-e^{2\varphi/3}\qty[\frac{1}{2}(\partial\varphi)^2\Box\varphi+\frac{1}{3}(\partial\varphi)^4],\nn\\
    e^{2\varphi/3}(\nabla\nabla\varphi)^2&\to e^{2\varphi/3}\qty[-R_{\alpha\beta}\partial^\alpha\varphi\partial^\beta\varphi+(\Box\varphi)^2+\frac{2}{3}(\partial\varphi)^2\Box\varphi-\frac{2}{3}\nabla_\alpha\nabla_\beta\varphi\partial^\alpha\varphi\partial^\beta\varphi],\nn\\
    e^{-2\varphi/3}G^2_{\alpha\beta}\nabla^\alpha\nabla^\beta\varphi&\to -\nabla^\alpha\qty(e^{-2\varphi/3}G_{\alpha\gamma})G_\beta{}^\gamma\partial^\beta\varphi+e^{-2\varphi/3}G^2\qty[\frac{1}{4}\Box\varphi-\frac{1}{6}(\partial\varphi)^2],\nn\\
    e^{10\varphi/3}H^2_{\alpha\beta}\nabla^\alpha\nabla^\beta\varphi&\to -\nabla^\alpha\qty(e^{10\varphi/3}H_{\alpha\gamma})H_\beta{}^\gamma\partial^\beta\varphi+e^{10\varphi/3}H^2\qty[\frac{1}{4}\Box\varphi+\frac{5}{6}(\partial\varphi)^2],\nn\\
    e^{4\varphi/3}G_{\alpha\gamma}H_{\beta}{}^\gamma\nabla^\alpha\nabla^\beta\varphi&\to -\frac{1}{2}e^{-2\varphi}\nabla^\alpha\qty(e^{10\varphi/3}H_{\alpha\gamma})G_\beta{}^\gamma\partial^\beta\varphi-\frac{1}{2}e^{2\varphi}\nabla^\alpha\qty(e^{-2\varphi/3}G_{\alpha\gamma})H_\beta{}^\gamma\partial^\beta\varphi\nn\\
    &\quad\ +e^{4\varphi/3}G\cdot H\qty[\frac{1}{4}\Box\varphi+\frac{1}{3}(\partial\varphi)^2].
\end{align}
We can write the four-derivative Einstein frame Lagrangian in the $C_{112}$ form using the field identification (\ref{eq:DWto112}).  After appropriate field redefinitions to remove the $\nabla_\alpha\nabla_\beta\varphi$ terms, we finally arrive at the expression (\ref{DWc2+16R2EinFinal}), where the CP-odd terms were rewritten using (\ref{eq:epsid}).

\section{Field redefinitions for the reduction from six dimensions}
\label{app:6dim}

The second derivative terms in the six-dimensional Lagrangian, (\ref{eq:sixdimact}), can be removed by use of the two-derivative equations of motion%
\footnote{Note that the metric equation of motion we have presented here is obtained by subtracting the dilaton equation of motion from the Einstein equation.}%
\begin{align}
    \mathcal E_{\hat g,MN}&=R_{MN}(\Omega)+2\hat\nabla_M\hat\nabla_N\phi-\frac{1}{4}\hat H^2_{MN},\nn\\
    \mathcal E_{\phi}&=R(\Omega)+4\Box\phi-4(\partial\phi)^2-\frac{1}{12}\hat H^2,\nn\\
    \mathcal E_{\hat B,MN}&=\hat\nabla^P\qty(e^{-2\phi} \hat H_{MNP}),
\end{align}
along with integration by parts and Bianchi identities. In particular, we can make the replacements
\begin{align}
    \qty(\hat\nabla\hat H)^2&\to 3R_{MNPQ}(\Omega)\hat H^{MNR}\hat H^{PQ}{}_R-\frac{3}{4}(\hat H_{MN}^2)^2,\nn\\
    \hat\nabla_M\hat\nabla_N\phi\partial^M\phi\partial^N\phi&\to-\frac{1}{2}(\partial\phi)^2\hat\Box\phi+(\partial\phi)^4,\nn\\
    \hat H^2_{MN}\hat\nabla^M\hat\nabla^N\phi&\to \frac{1}{6}\hat H^2\hat\Box\phi-\frac{1}{3}\hat H^2(\partial\phi)^2,\nn\\
    \qty(\hat\nabla_M\hat\nabla_N\phi)^2&\to-4(\partial\phi)^2\hat\Box\phi+4(\partial\phi)^4+\qty(\hat\Box\phi)^2-\frac{1}{4}H^2_{MN}\partial^M\phi\partial^N\phi,\nn\\
    \hat H^{MNP}\partial^Q\phi\hat \nabla_Q \hat H_{MNP}&\to-\frac{1}{2}\hat H^2\hat\Box\phi+\hat H^2(\partial\phi)^2,\nn\\
    \hat\Box\phi&\to 2(\partial\phi)^2-\frac{1}{12}\hat H^2,
\end{align}
and
\begin{equation}
    \epsilon^{MNPQRS}\hat H_{QRS}\hat H_{MN}{}^T\hat \nabla_P\hat \nabla_T\phi\to 0.
\end{equation}
The resulting Lagrangian takes on a more streamlined form, which is given in (\ref{eq:6dAction}).

The structure of the six-dimensional superinvariant can be highlighted by decomposing the three-form field strength $\hat H$ into its self-dual and anti-self-dual components.  However, instead of directly substituting $\hat H=\hat H^{(+)}+\hat H^{(-)}$ into (\ref{eq:6dAction}), we may instead make use of the identity
\begin{align}
    -\hat H_{MN}^2\partial^M\phi\partial^N\phi+\frac{1}{6}\epsilon^{MNPQRS}\hat H_{QRS}\hat H^T{}_{NP}\partial_M\phi\partial_T\phi=-2(\hat H^{(-)})^2_{MN}\partial^M\phi\partial^N\phi-\frac{1}{6}\hat H^2(\partial\phi)^2,
\end{align}
where we have used
\begin{equation}
    \hat H^{(\pm)}=\frac{1}{2}\qty(\hat H\pm\star\hat H),\qquad (\star\hat H)^{MNP}=\frac{1}{3!}\epsilon^{MNPQRS}\hat H_{QRS},
\end{equation}
and the identity~\cite{Liu:2019ses}
\begin{equation}
    \hat H_{(M}^{(+)}{}^{AB}\hat H_{N)AB}^{(-)}=\frac{1}{12}\hat H^2 g_{MN}.
\end{equation}
In addition, using \eqref{eq:SezginIdentity}, we can write
\begin{equation}
    \frac{1}{144}\epsilon^{MNPQRS}\hat H_{QRS}\hat H^2_{MT}\hat H_{NP}{}^T=-\frac{1}{24}(\star\hat H)^{MNP}(\hat H^3)_{MNP}+\frac{1}{16}(\star\hat H)_{MPQ}\hat H_N{}^{PQ}\hat H^2_{MN}.
\end{equation}
This then means that the $\hat H^4$ type terms combine as
\begin{equation}
    \frac{1}{24}\hat H^4-\frac{1}{24}(\star\hat H)^{MNP}(\hat H^3)_{MNP}=\frac{1}{12}\hat H^{(-)}_{MNP}\hat H^3_{MNP},
\end{equation}
and the $(\hat H_{MN}^2)^2$ type terms combine as
\begin{equation}
    -\frac{1}{16}(\hat H_{MN}^2)^2+\frac{1}{16}(\star\hat H)_{MPQ}\hat H_N{}^{PQ}\hat H^2_{MN}=-\frac{1}{8}\hat H^{(-)}_{MPQ} \hat H_N{}^{PQ}\hat H^2_{MN}.
\end{equation}
We also have
\begin{equation}
    \frac{1}{12}\hat H^{(-)}_{MNP}\hat H^3_{MNP}-\frac{1}{8}\hat H^{(-)}_{MPQ}\hat H_N{}^{PQ}\hat H^2_{MN}=-\frac{1}{6}(\hat H^{(-)})^4-\frac{1}{96}(\hat H^2)^2.
\end{equation}
Making use of these identities, we then end up with the quartic tensor multiplet coupling shown in (\ref{eq:6dactionCorrected}).


\bibliographystyle{JHEP}
\bibliography{cite}

\end{document}